\documentclass[journal]{IEEEtran}
\usepackage{graphicx}
\usepackage{amssymb}
\usepackage{amsmath}
\usepackage{algorithmic}
\usepackage{textcomp}
\usepackage{algorithm}
\usepackage{subcaption}
\usepackage{multirow}
\usepackage{color}
\usepackage{booktabs}
\ifCLASSINFOpdf
\else
\fi

\begin{document}

\title{STC-IDS: Spatial-Temporal Correlation Feature Analyzing based Intrusion Detection System for Intelligent Connected Vehicles}

\author{Pengzhou~Cheng,
		Mu~Han*,~\IEEEmembership{Member,~IEEE},
		Aoxue Li,
       and~Fengwei~Zhang
        
\thanks{This research was supported by the following funds: Natural Science Fund for Colleges and Universities in Jiangsu Province [12KJD580002]; the Jiangsu Graduate Innovation Fund [KYLX1057]; the Key Research and Development Plan of Jiangsu Province in 2017 (Industry Foresight and Generic Key Technology) [BE2017035]. (Corresponding author: Mu Han.)}

\thanks{Pengzhou Cheng is with the Department of Electronic Information and Electrical Engineering, Shanghai Jiao Tong University, Shanghai, 201100, China (e-mail: pengzhouchengai@gmail.com).}

\thanks{Mu Han is with the School of Computer Science and Communication Engineering and Jiangsu Key Laboratory of Security Technology for Industrial Cyberspace
	Jiangsu University, Zhenjiang 212013, China (e-mail: hanmu@ujs.edu.cn).}

\thanks{Aoxue Li is with the School of Automotive and Traffic Engineering, Jiangsu University, Zhenjiang 212013, China (e-mail: ujsliaoxue748@163.com).}
\thanks{Fengwei Zhang is with the Department of Computer Science and Engineering, Southern University of Science and Technology, Shenzhen 518000, China (e-mail:zhangfw@sustech.edu.cn).}
}

\markboth{}%
{Shell \MakeLowercase{\textit{et al.}}: Bare Demo of IEEEtran.cls for IEEE Journals}

\maketitle

\begin{abstract}
Intrusion detection is an important defensive measure for automotive communications security. Accurate frame detection models assist vehicles to avoid malicious attacks. Uncertainty and diversity regarding attack methods make this task challenging. However, the existing works have the limitation of only considering local features or the weak feature mapping of multi-features. To address these limitations, we present a novel model for automotive intrusion detection by spatial-temporal correlation features of in-vehicle communication traffic (STC-IDS). Specifically, the proposed model exploits an encoding-detection architecture. In the encoder part, spatial and temporal relations are encoded simultaneously. To strengthen the relationship between features, the attention-based convolutional network still captures spatial and channel features to increase the receptive field, while attention-LSTM builds meaningful relationships from previous time series or crucial bytes. The encoded information is then passed to detector for generating forceful spatial-temporal attention features and enabling anomaly classification. In particular, single-frame and multi-frame models are constructed to present different advantages respectively. Under automatic hyper-parameter selection based on Bayesian optimization, the model is trained to attain the best performance. Extensive empirical studies based on a real-world vehicle attack dataset demonstrate that STC-IDS has outperformed baseline methods and obtains fewer false-alarm rates while maintaining efficiency.
\end{abstract}


\IEEEpeerreviewmaketitle

\section{Introduction} \label{sec1}

Nowadays, a large number of electronic control units (ECUs) have replaced the mechanical control units to manage the assorted functions of in-vehicle control systems. The ECUs are interconnected to exchange varied vehicle information with each other via networks referred to as in-vehicle networks (IVNs) such as controller area networks (CAN) \cite{song2021self}. Along with local interconnected network LIN and FlexRay, CAN is well-known and most employed as the de-factor standard for IVNs \cite{kim2008gateway, hoppe2011security}. It is noteworthy that CAN was developed as a broadcast-based communication protocol that supports the maximum baud rate up to 1Mb per-second. Furthermore, the fault-tolerant detection mechanism guarantees the stability of message transmission.

However, the CAN bus is potentially vulnerable to attacks owing to the lack of security mechanisms such as encryption, access control, and authentication \cite{miller2015remote, li2020efficient, gao2020mas}. Cyber security is becoming a major concern for IVNs systems as increasingly more security researchers demonstrate their ability to launch attacks on actual vehicles \cite{liu2020eurus}. What can be investigated is various attacks have been threatening several significant components of IVNs \cite{miller2015remote,checkoway2011comprehensive,petit2014potential}. For instance, 360 cyberattack Lab adopted electronic radio-frequency technology to successfully hack into Tesla in 2015 \cite{aliwa2021cyberattacks}. Miller \textit{et al}. invaded the Jeep Cherokee's IVNs system using an open Wi-Fi port and reprogrammed the firmware of ECU. They succeeded in taking control of a wide range of vehicle functions (e.g., disabling the brakes and stopping the engine), triggering a recall of 1.4 million vehicles \cite{miller2015remote}. Thereafter, the electric features lift, warning lights, airbag, and tire pressure monitoring system (TPMS) have also become the target of attack \cite{rouf2010security}. Incredibly, these attacks have multiple ways of being performed. As such, the study on the security of IVNs is attracting significant attention from security researchers \cite{yan2020secure, chen2021camdar}.

There are many available methods that have the ability to protect IVNs secure, where IDS as an effective defense method has attracted more attention from researchers \cite{wu2019survey, tianqing2021resource}. Currently, a host of IDS schemes are rule-based and statistical-based. Although accuracy and efficiency are excellent about some common attacks, the passive characteristic and the constant need for updates result in a certain restriction \cite{he2013effective}. With the increase in vehicle computing power and the maturity in machine learning (ML) technology, they promote the further development of IDS \cite{cai2021capbad}. Real-time, higher detection, and lower false-positive rates have been a fervent research problem for deep learning-based in-vehicle IDS. Simultaneously, inadequate feature extraction, complex network structure, and more parameters, also are pending breakthroughs \cite{song2021self, mo2022attacking}. To address such limitations, variants of IDS based on deep learning have been proposed in recent years \cite{song2021self,qin2021application,olufowobi2019saiducant}. However, these variants merely consider the partial features, either time-series of CAN ID or CAN data field. Moreover, spatial-temporal correlation features have been shown to better capture the details of the message in anomaly detection \cite{tariq2019detecting,zhang2019deep,sun2021anomaly}, whereas how to create represent the relationship between spatial-temporal features becomes a vital concern to elevate detection performance in the automotive intrusion \cite{kuang2019deepwaf, cai2021appm}.

Thus, the purpose of this paper is to provide STC-IDS (Spatial-Temporal Correlation Features Analyzing based-enhanced Intrusion Detection System). Based on an encoding-detection architecture, the intuition first trained a boosted convolutional LSTM parallel feature extraction model. Improved attention-based LSTM network (A-LSTM) captures the temporal features and builds important relationships from previous sequences or crucial bytes. Meanwhile, a reduced VGGNet network can learn the spatial features from CAN frames. Moreover, the attention convolutional block (A-Conv2D) enables attaining a broad perspective through multi-channel features to refine feature mapping.Unlike many previous methods, it can concentrate more on changes in crucial areas and ignore bytes that are regular and unchanging. Afterward, spatial-temporal correlations between features are established to feed into the detector as a two-class classification problem. Note that both single-frame and multi-frame models are considered in this paper to present different advantages respectively.

This paper makes the following contributions.
\begin{enumerate}
	\item Based on analyzing spatial-temporal correlation features of CAN messages in detail, we propose an enhanced convolutional LSTM spatial-temporal feature encoder with attention. The single-frame model automatically captures the important byte relationships of each CAN frame and uses convolutional components to find where key bytes are. Moreover, the multi-frame model captures significant relationships from previous time series, in which the attention convolutional block, assisted by spatial attention and channel attention, is able to snap changes in crucial areas.
	\item Further, the detector achieves anomaly detection for the constructed representative spatial-temporal correlation features after multi-view learning. We evaluated the detection performance of our scheme using a publicly available real vehicle CAN dataset. We also compare it with the baseline model and show a significant improvement in detection performance and a reduction in false-positive rates and error rates.
	\item By performing the injection attack in the same way, we calculated the detection efficiency of the model on real vehicles. The multi-frame model has sufficient ability to satisfy real-time detection. In addition, the single-frame model combined with database retrieval has the ability to trace anomalous ECUs.
\end{enumerate}

The rest of this paper is organized as follows. Section \ref{sec2}  presents a discussion of related work on CAN-based IDS. Section \ref{sec3} introduces the CAN bus protocol, vulnerability, and analyzes the CAN frame with spatial-temporal. Section \ref{sec4} proposes the parallel network model based on spatial-temporal features analysis. The experiment result of the proposed model on the public real-vehicle dataset is described in Section \ref{sec5}. Finally, we conclude this study and look ahead at in-vehicle IDS potential perspectives.

\section{Related Work}\label{sec2}
In this section, we provide an in-depth discussion about the research situation in anomaly detection and intrusion detection for in-vehicle CAN communication systems. They are divided into four categories, namely specification-based, fingerprint-based, statistical-based and machine learning-based approaches, as summarized in Table \ref{tab1}.

\begin{table*}[t]%
	\centering
	\caption{Comparison of in-vehicle intrusion detection systems.\label{tab1}}%
	\resizebox{\textwidth}{!}{
		\begin{tabular}{cclcll}%
			\toprule
			\textbf{Categories} & \textbf{Research work} &\textbf{Technique} & \textbf{Evaluation Data} & \textbf{Contribution} & \textbf{Limitation} \\
			\midrule
			\multicolumn{1}{c}{\multirow{3}{*}{\begin{tabular}[c]{@{}c@{}}Specification \\ based\end{tabular}}} & Studnia \textit{et al}. \cite{studnia2018language}                    & \begin{tabular}[l]{@{}l@{}}a list of signatures derived from\\ CAN data set \end{tabular}                                              & real                                  & \begin{tabular}[l]{@{}l@{}}automatically generating a set of \\forbidden sequences\end{tabular}     & \begin{tabular}[l]{@{}l@{}}the length of CAN bus words may\\ not be known a priori \end{tabular}                                                         \\ \cmidrule{2-6} \multicolumn{1}{c}{} 
			& Dagan \textit{et al}. \cite{dagan2016parrot}                     & \begin{tabular}[l]{@{}l@{}}an anti-spoofing system detects \\ CAN message IDs by each ECU \\ that are not sent by the ECU itself \end{tabular}                                                                                                                                          & \multicolumn{1}{c}{\begin{tabular}[l]{@{}c@{}}real \\ and simulation\end{tabular}}                   & \begin{tabular}[l]{@{}l@{}}effective resistance to spoofing \\ attacks\end{tabular}                                    &\begin{tabular}[l]{@{}l@{}}each ECU takes on the IDS role,\\ adding a certain burden to com-\\munication\end{tabular}                            \\ \cmidrule{2-6} \multicolumn{1}{c}{} 
			& Olufowobi \textit{et al}. \cite{olufowobi2019saiducant} & \begin{tabular}[l]{@{}l@{}}worst-case response time analysis \end{tabular}                                                                                                              & \begin{tabular}[l]{@{}c@{}}real \\ and synthesized\end{tabular} & \begin{tabular}[l]{@{}l@{}}real-time parameter estimation al-\\gorithm developed in a black-box\\ approach \end{tabular}      & \begin{tabular}[l]{@{}l@{}}high false positive rate and relatively\\ low performance on open dataset \end{tabular}\\ \cmidrule{1-6} 
			
			\multicolumn{1}{c}{\multirow{2}{*}{\begin{tabular}[c]{@{}c@{}}Fingerprint \\ based\end{tabular}}}   & Cho and Shin \cite{cho2016fingerprinting}                      & \begin{tabular}[l]{@{}l@{}}analyzed the clock skew of ECUs \end{tabular}                                                                                                                                  & \begin{tabular}[l]{@{}c@{}}real \\ and simulation\end{tabular}                   & \begin{tabular}[l]{@{}l@{}}proposed an ECU pro-filing meth-\\od according to electrical signal\\ characteristics \end{tabular}               & \begin{tabular}[l]{@{}l@{}}workable only to periodic messages\\ excluding non-periodic messages \end{tabular}                                            \\ \cmidrule{2-6}  \multicolumn{1}{c}{} 
			& Choi \textit{et al}. \cite{choi2018voltageids}                      & \begin{tabular}[l]{@{}l@{}}established the electrical signal \\features of each ECU based on\\ time domain and frequency domain\end{tabular} & \begin{tabular}[l]{@{}c@{}}real \\and simulation\end{tabular}                    & \begin{tabular}[l]{@{}l@{}} 
				effective of making a distinction \\between ECU malfunctions and\\ a bus-off attack \end{tabular}                                              & \begin{tabular}[l]{@{}l@{}}the electrical properties of the veh-\\icle may change as it ages, necessit-\\ating correction.\end{tabular} \\ \midrule
			\multirow{2}{*}{\begin{tabular}[l]{@{}c@{}}Statistics \\ based\end{tabular}}   & Song \textit{et al}. \cite{song2016intrusion}                      & \begin{tabular}[l]{@{}l@{}}analyzed the time-intervals of the \\ CAN messages \end{tabular}                                                                                                 & \begin{tabular}[l]{@{}c@{}}real \\ and synthesized\end{tabular}                  & \begin{tabular}[l]{@{}l@{}}offered effectiveness of the me-\\thod for injection attack detection \end{tabular}                             & \begin{tabular}[l]{@{}l@{}}workable only to periodic messages\\ excluding non-periodic messages \end{tabular}                                            \\ \cmidrule{2-6} 
			& Young \textit{et al}. \cite{young2019survey}                      & \begin{tabular}[l]{@{}l@{}}analyzed the time-intervals in the \\frequency domain \end{tabular}                                                                                                              & real                                  & \begin{tabular}[l]{@{}l@{}}improved frequency-based inject-\\ion attack detection method \end{tabular}                                  & \begin{tabular}[l]{@{}l@{}}high false positive rate, and only \\ identify injection attack \end{tabular}                                                  \\ \midrule
			\multirow{8}{*}{\begin{tabular}[l]{@{}c@{}}ML \& DL \\ based\end{tabular}}      
			& Kang \textit{et al}. \cite{kang2016intrusion}                       & deep belief network (DBN)                                                                                                                                              & simulation                            & \begin{tabular}[l]{@{}l@{}}presented unsupervised statistical\\ feature extraction algorithms for\\ ECU messages  \end{tabular}                                                    & \begin{tabular}[l]{@{}l@{}}extremely time-consuming in the\\ training phase, evaluation only based\\ on simulation data \end{tabular}                                                 \\ \cmidrule{2-6} 
			& Yang \textit{et al}. \cite{yang2021mth}                       & \begin{tabular}[l]{@{}l@{}}signature-based and anomaly-based\\ multi-layer hybrid IDS   \end{tabular}                                                                                                                                           & real                            & \begin{tabular}[l]{@{}l@{}}proposed an IDS that is effective\\ for attacks on both in-vehicle and\\ external networks  \end{tabular}                                                    & \begin{tabular}[l]{@{}l@{}}spatial-temporal features and important\\ regional features are ignored \end{tabular}                                                 \\ \cmidrule{2-6}
			& Tariq \textit{et al}. \cite{tariq2020cantransfer}                      & \begin{tabular}[l]{@{}l@{}}convolutional LSTM Network \\ and transfer learning                  \end{tabular}                                                                                                                            & real                                  & \begin{tabular}[l/]{@{}l@{}}improved performance of unknown\\ attack detection \end{tabular}                                              & \begin{tabular}[l]{@{}l@{}}known attacks have poor relative de-\\tection performance \end{tabular}                                                        \\ \cmidrule{2-6} 
			& Pawelec \textit{et al}. \cite{pawelec2019towards}   &\begin{tabular}[l]{@{}l@{}} LSTM prediction at the bit level\end{tabular}   & real  & \begin{tabular}[l]{@{}l@{}}avoided reverse engineering\\ proprietary encodings \end{tabular}    & \begin{tabular}[l]{@{}l@{}}limited attack detection range \end{tabular} \\ \cmidrule{2-6} 
			& Qin \textit{et al}. \cite{qin2021application}                        & \begin{tabular}[l]{@{}l@{}}LSTM network and five loss\\ function \end{tabular}                                                                                                                                                            & real                                  & \begin{tabular}[l]{@{}l@{}}presented two predicted model on\\ different data format \end{tabular}                                                       & \begin{tabular}[l]{@{}l@{}}relatively poor detection accuracy          \end{tabular}                                                                  \\ \cmidrule{2-6} 
			& Song \textit{et al}. \cite{song2020vehicle}                       & deep CNN                                                                                                                                                         & real                                  & \begin{tabular}[l]{@{}l@{}}reduced Inception-resent model\\ complexity to adapt IDS and im-\\prove detection performance\end{tabular} & \begin{tabular}[l]{@{}l@{}} the spatial-temporal relationship \\ of CAN frames is ignored \end{tabular}                  \\                               
			\cmidrule{2-6} 
			& STC-IDS (Ours)                       & \begin{tabular}[l]{@{}l@{}}parallel convolutional LSTM\\ attention network \end{tabular}                                                                                                                                                         & real                                  & \begin{tabular}[l]{@{}l@{}} important concerns for spatial-\\temporal features and crucial \\ features
			\end{tabular} & \begin{tabular}[l]{@{}l@{}} more unknown attacks are ignored \end{tabular}                                                  \\ \bottomrule
			
	\end{tabular}}
\end{table*}

\subsection{Intrusion detection model based on specification}
The specification-based IDS focuses on defining system specifications, such as protocols and frame formats. When packets mismatch the system specification, an exception alarm is raised. In 2016, Dagan \textit{et al}. \cite{dagan2016parrot} introduced an anti-spoofing system that detects malicious messages using each ECU, i.e., by detecting CAN message ID that was not sent by the ECU itself. Thereafter, the ECU informs the IDS, and then an interrupt pulse is sent to the CAN bus to overwrite the spoofed message. However, each ECU undertakes the IDS role, which increases a certain burden on communication.

In 2018, Studnia \textit{et al}. \cite{studnia2018language} presented a signature-based IDS which utilize a list of signature derived from CAN dataset. However, this method is subject to the limitation that the length of the CAN bus words may not be known a priori. Recently, Olufowobi \textit{et al}. \cite{olufowobi2019saiducant} proposed a real-time IDS based on specification. The algorithm extracted the timing model to detect anomalies through observing CAN traffic rather than depending on predefined specifications, yet exhibited relatively poor performance in the real attack dataset.

\subsection{Intrusion detection model based on fingerprint}
The fingerprint-based approaches are mainly based on profiles defined by ECU characteristics to implement anomaly detection. In 2016, Cho and Shin \cite{cho2016fingerprinting} proposed clock-based IDS to analyze ECUs periodic frequency. The method established the ECUs clock baseline through the recursive least squares algorithm (RLS) to detect intrusion. But it workable only to periodic messages excluding non-periodic messages. 

Interestingly, Cho \textit{et al}. \cite{choi2018voltageids} found a method to establish the electrical signal characteristics of each ECU using the physical layer data of CAN communication, and harness these signal characteristics as the fingerprint for each ECU. Regrettably, the electrical characteristics may change as the vehicle ages, and thus the IDS needs to keep updating.

\subsection{Intrusion detection model based on statistical}
Unlike previous methods, the statistics-based approach implements anomaly detection by means of statistical information obtained from CAN traffic at the network level. Song \textit{et al}. \cite{song2016intrusion} introduced a lightweight IDS in 2016 that detected anomalies by monitoring the abnormally shortened intervals between messages. Although the proposed algorithm could have highly sensitive to common injection attacks and low computing cost, it cannot detect irregular incoming messages. 

Furthermore, Young \textit{et al}. \cite{young2019survey} comprehensively analyzed the frequency characteristics of CAN messages in various driving modes, such as reverse, acceleration and hold speed, and then proposed a frequency-based intrusion detection system. In spite of the high detection accuracy, there is a high false alarm rate. Manifestly, an IDS based on conditional statistical relationship analysis to learn the normal behavior of the system can detect manipulations and incorrect payload values \cite{tomlinson2018towards}, but still does not satisfy the high detection rate and low latency required by present-day IVNs for anomalous traffic \cite{zhang2021delay}.

\subsection{Intrusion detection model based on deep learning}
Machine learning (ML) and Deep learning (DL) based intrusion detection systems are an excellent option for extracting and learning normal or abnormal behavior, which provide models with detect and predict ability \cite{ai2021adversarial, hu2021mhat}. Kang \textit{et al}. \cite{kang2016intrusion} constructed a deep confidence network under unsupervised learning to detect if anomalies deviate from normal. However, inefficient and only validation of simulation data is not sufficient. 

In 2019, Pawelec \textit{et al}. \cite{pawelec2019towards} proposed a 3-layer LSTM neural network to predict the data payload for each CAN ID, which avoided reverse engineering proprietary encoding. Similarly, Qin \textit{et al}. \cite{qin2021application} also implemented anomaly detection for CAN bus based on timing features by LSTM and re-considered two data formats of CAN frames. Although these methods are implemented at the CAN bits level, they only consider timing characteristics and have relatively poor detection performance. Recently, convolutional neural network (CNN) have been implemented for traffic detection and praised for their high detection efficiency \cite{wang2017malware}. 

Song \textit{et al}. \cite{song2020vehicle} presented a reduced inception residual network to construct an IDS capable of detecting spoofing and denial of service (DoS) attacks in a continuous pattern of vehicular traffic. Since the assistance of spatial-temporal relation is not taken into account, there is still room for improvement in the false positive rate. In other studies, Tariq \textit{et al}. \cite{tariq2020cantransfer} introduced a convolutional LSTM-based intrusion detection method. Although it displayed excellent detection performance in unknown attack than transfer learning, known attacks performance relatively worse that may be caused by the relevance of the CAN message features being discarded. Moreover, a multi-tiered hybrid IDS that incorporates a signature-based IDS and an anomaly-based IDS is proposed to detect both known and unknown attacks on vehicular networks by Yang \textit{et al}. \cite{yang2021mth} in 2021. Their model has been proven that is effective for attacks on both in-vehicle and external networks. However, modeling with spatial-temporal features might take performance a step further.

However, these ML\&DL-based methods have different in selecting the detection domain, typically the detection arbitration domain, the detection data domain, and the similar to our work that the spatial-temporal feature extraction. In a nutshell, traditional methods based on specification, fingerprint, and statistical, have limitations in terms of reliance on anomaly feature libraries, message frequency, message time domain, and fingerprint information. Instead, it is imperative in the ML\&DL area to improve automotive IDS detection performance and reduce false positives by complementing spatial-temporal features in a limited message communication mode. Furthermore, the model trained with limited spatial-temporal features will not effectively improve the detection performance. 

Considering the nature of attention mechanisms to capture essential features, the generation of spatial-temporal attention features is one of the potential ways to address the above problem \cite{jiang2022gatrust}. Hence, modeling the normal behavior of CAN packets in combination with spatial-temporal attention features and then discovering the difference between anomalies and target traffic is still one of the ways to improve in-vehicle IDS.

\section{AN IN-DEPTH OVERVIEW OF CAN BUS DATASET}\label{sec3}
\subsection{Vulnerabilities of in-vehicle networks}
\begin{figure}[t]
	\centering	
	\includegraphics[width=1\linewidth]{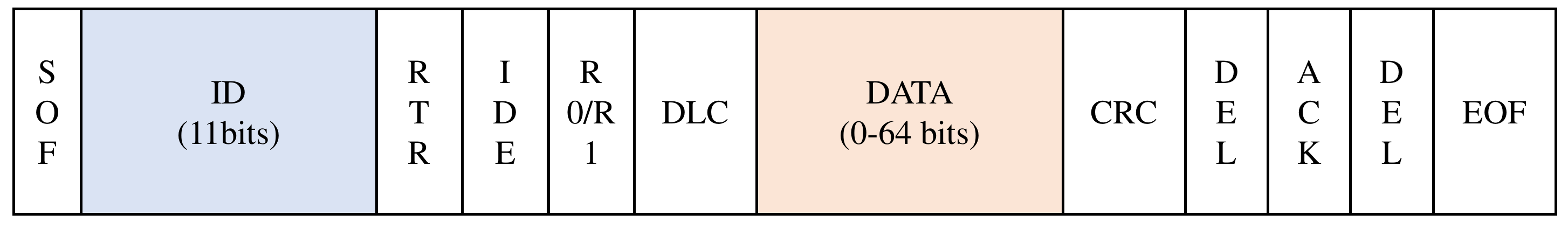}
	\caption{Structure of a CAN 2.0A message frame.}
	\label{Fig.1}
\end{figure}

Intelligent Connected Vehicles (ICVs), integrating modern computing and communication technologies, is designed to improve user experience and driving safety. As the most significant communication medium, the CAN is the most prevalent bus topology network employed in contemporary vehicles owing to its low cost and complexity, high reliability, and fault-tolerance characteristics \cite{yang2019tree, liu2020encodeore}. All ECUs, connected to the CAN bus, are capable of exchanging messages in the form of data frames \cite{jadhav2015review, cai2022sarm}. Fig \ref{Fig.1} presents the structure of a CAN message, consisting of seven fields \cite{al2019intrusion}: 1) start of frame (1 bit); 2) arbitration field (12 bits for standard frames, 29 bits for extended frames); 3) control field (6 bits); 4) data field (Maximally 8 bytes); 5) cyclic redundancy code (CRC) field; 6) acknowledge (ACK) field; and 7) end of frame.

\begin{figure}[t]
	\centering	
	\includegraphics[width=1\linewidth]{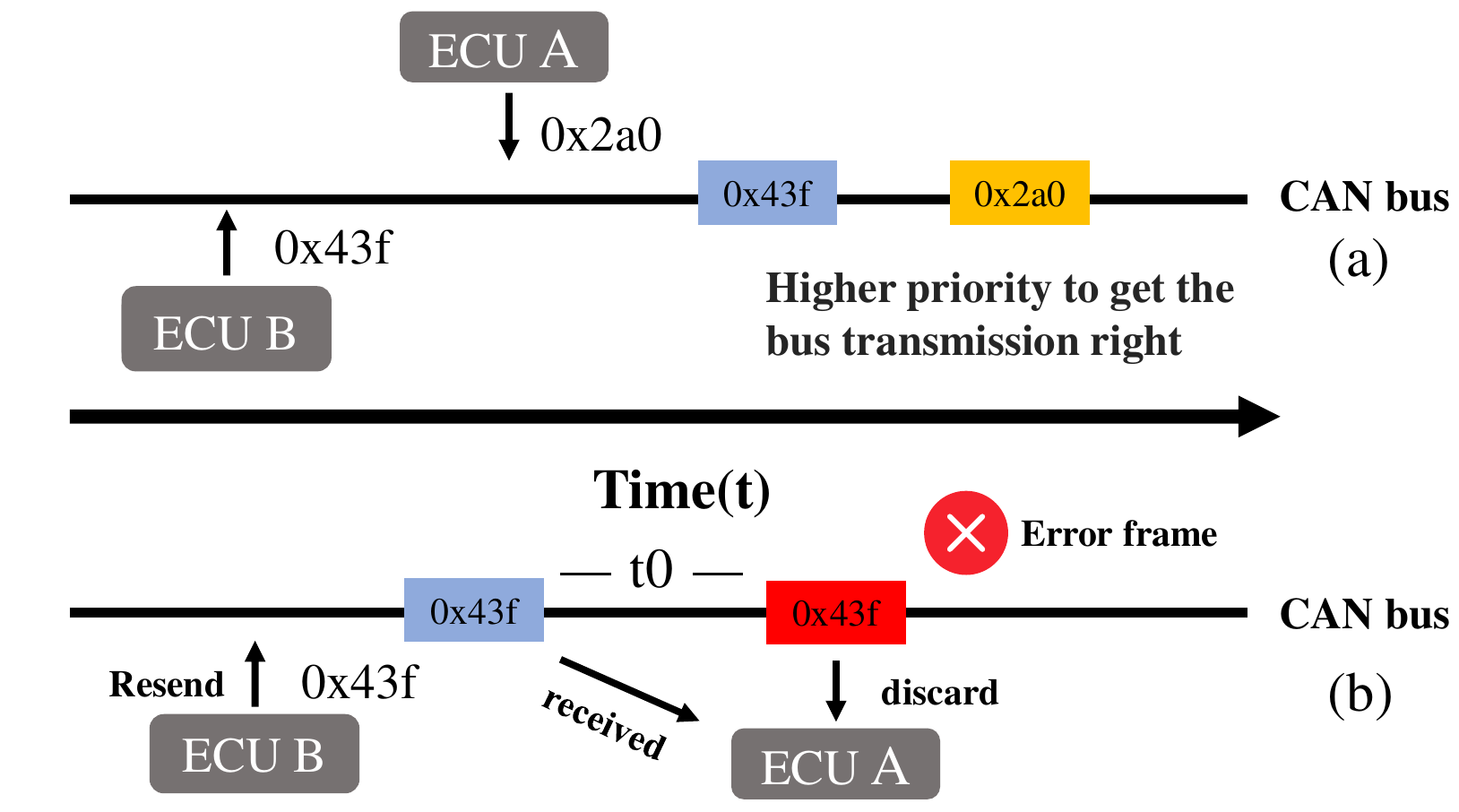}
	\caption{Conceptual diagram of the message priority and CRC detection.}
	\label{Fig.2}
\end{figure}

In the entire CAN frame, the most important is the arbitration field and the data field, as the arbitration field determines the priority of the message \cite{davis2007controller}, shown in Fig \ref{Fig.2}(a); the data field contains the actual transmitted data that defines the node actions. Moreover, if an error is detected by CRC field, the receiving node will discard the received error message, while the sending node will only assume a transient fault on the bus and enter arbitration to resend the message frame \cite{tindell1995calculating,punnekkat2000response}, shown in Fig \ref{Fig.2}(b). To guarantee the system consistency, the ECU broadcasts messages at regular intervals in spite of the data values have not changed. However, security problems were poor-needy thought out at the beginning of the CAN bus design \cite{bar2020Intrusion}, including its broadcast transmission strategy, lack of authentication and encryption, and unsecured priority scheme. Hence, many network traffic injection attacks are possible. 

\begin{figure}[t]
	\centering	
	\includegraphics[width=1\linewidth]{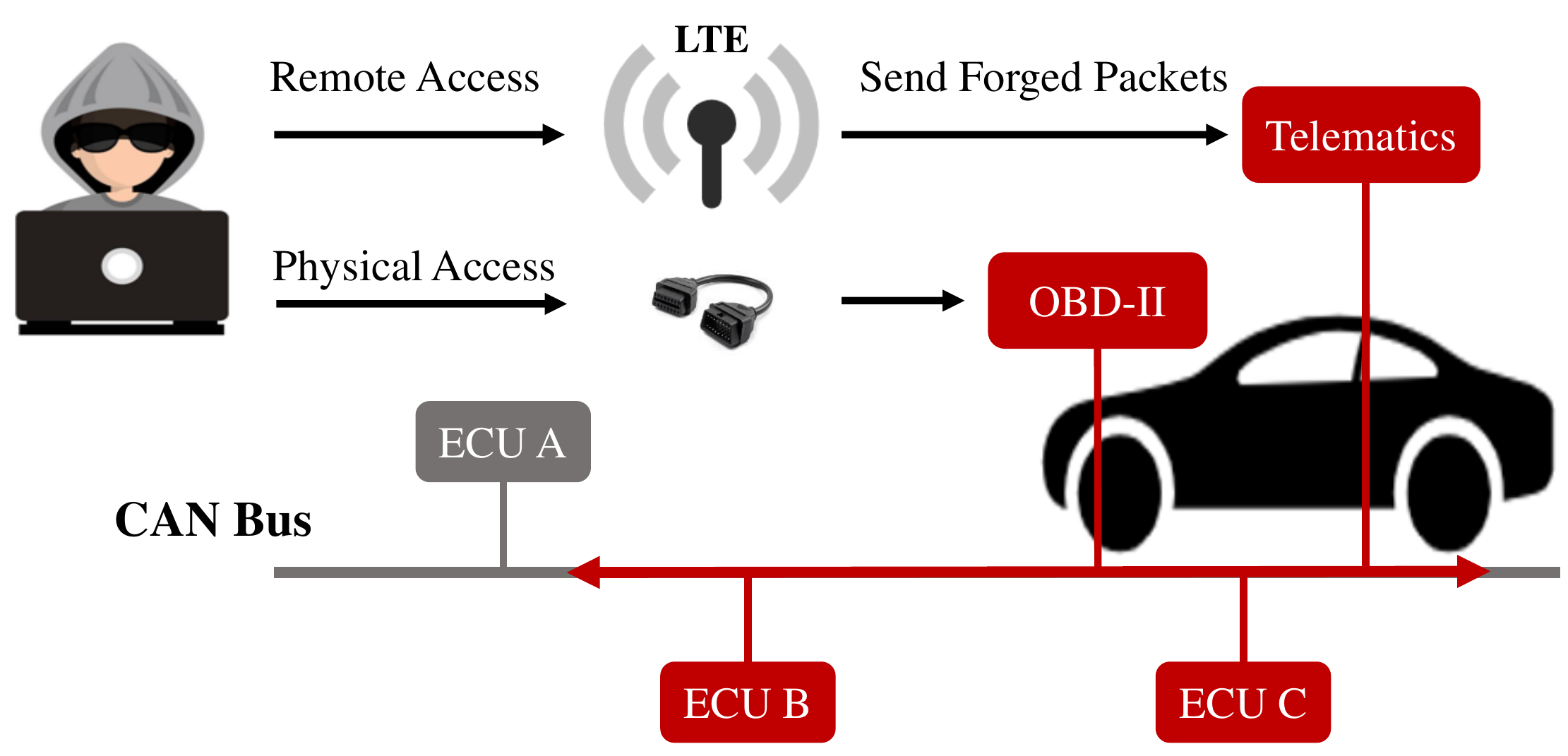}
	\caption{The scenario in which an attacker performs an injection attack can be a remote attack or a physical attack can be implemented.}
	\label{Fig.3}
\end{figure}

This directly motivates the adversary to attack in-vehicle networks in a variety of ways, as shown in Fig \ref{Fig.3}. Clearly, the adversary can not only through the OBD-II port for physical attacks but also implement a remote attack easily (e.g., Wi-Fi or Bluetooth) \cite{aliwa2021cyberattacks}. Types of such attacks include flooding the bus with messages designed to circumvent legitimate messages or using spoofed bus identifiers with invalid information \cite{song2020vehicle}. Furthermore, there are more sophisticated and stealthy attacks \cite{GIDS}. These attacks appear to be legitimate traffic sequences that are tough to distinguish from normal messages \cite{song2021self}.

Once the attacker has successfully compromised, it has the opportunity to forge ECU nodes and take control of incumbent nodes to inject nefarious messages \cite{li2020dynamic, ai2021csrt}. In the CAN protocol, the connected nodes are synchronized with the current vehicle state by accepting the data field bits of the frame \cite{song2020vehicle}. Consequently, in order to successfully deceive the ECU, the adversary must insert tampering messages in a high frequency and priority manner by following the target CAN ID message immediately after the message \cite{cai2022deep}. If the attacker injects a high-priority CAN frame, where the data field is populated with a status command to turn off the ``wiper", the driver loses judgment and even serious traffic accidents in rainy conditions while driving at high speed \cite{yan2021ppcl}.

\subsection{Spatial-temporal correlation feature analyzing for CAN dataset}
\begin{figure}[t]
	\centering	
	\includegraphics[width=1\linewidth]{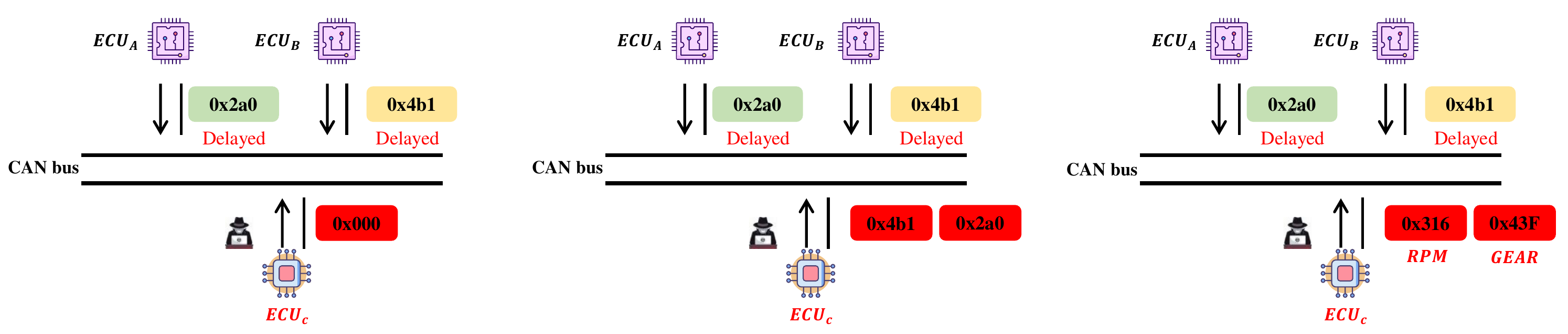}
	\caption{Illustration of the injection process of utilized in-vehicle intrusion dataset.}
	\label{Fig.4}
\end{figure}

In this paper, we utilized car-hacking dataset which is published by Song \textit{et al}. \cite{GIDS}. This data set is injected with four attacks, respectively DoS attack, fuzzy attack, spoofing attack including RPM and GEAR, as illustrated in Fig \ref{Fig.4}. The detailed injection rules are as follows.
\begin{enumerate}
	\item DoS attack: DoS attacks in the dataset are to inject a high priority message with a `0x000' CAN ID every 0.3 milliseconds, with the data field populated with 0.
	\item Fuzzy attack: Fuzzy attacks in dataset are injected every 0.5 milliseconds with CAN messages where the CAN ID and DATA values are forged randomly.
	\item Spoofing attack: Spoofing attacks in dataset are injected messages every 1 millisecond with a specific CAN ID, e.g., related to RPM/Gear.
\end{enumerate}

Table \ref{table2} indicates the number of normal and injected messages in each attack dataset. In order to summarize the spatial-temporal details of normal CAN bus traffic during the operation of a real vehicle, we first investigated the communication patterns of different CAN IDs. Since the dataset is not publicly available as to the receiving sender and receiver of the data, the paper analyzes the CAN protocol table of a brand from our laboratory. We find that an ECU has a fixed set of CAN IDs (e.g. EMS, containing 0x101, 0x278, 0x281) and that the recipients of the different CAN IDs are also fixed (e.g. 0x101, containing TCU, ESP, EPB, T-BOX). This is the initial purpose to design a single-frame detection model capable of tracking unauthorized ECUs and protecting non-attacked ECUs.

\begin{table}[t]
	\centering
	\normalsize
	\caption{Overview of car-hacking dataset}
	\begin{tabular}{lcc}%
		\toprule
		\textbf{Attack   type} & \textbf{Normal   messages} & \textbf{Injected   messages} \\
		\midrule
		DoS   attack   & 3078250           & 587521              \\ 
		Fuzzy   attack & 3347013           & 491847              \\ 
		RPM   attack   & 2766522           & 597252              \\ 
		Gear   attack  & 2290185           & 654897              \\ 
		\bottomrule
	\end{tabular}
	\label{table2}
\end{table}

\begin{table}[t]
	\centering
	\caption{Partial ECU transmission, reception and time cycle of an automotive brand}
	\resizebox{\linewidth}{!}{
		\renewcommand{\arraystretch}{1.1}
	\begin{tabular}{lccl}%
		\toprule
		\textbf{ID}    & \textbf{Transmitter} & \textbf{periods} & \textbf{Receiver}     \\ 
		\midrule
		0x101 & EMS         & 10      & TCU,   ESP, EPB, T-BOX                          \\
		0x278 & EMS         & 10      & TCU,   GSM, ABS, ESP, EPS, EPB, PEPS \\
		0x281 & EMS         & 100     & TCU,   AC, ICU, HUD, T-BOX                      \\
		0x1A0 & TCU         & 10      & EMS,   GSM, ESP, EPB, PEPS, DRC, PDC            \\
		0x211 & ESP         & 10      & TCU,   ESP, EPB, T-BOX                          \\
		...   & ...         & ...     & ...                                             \\
		0x2EA & ABS         & 20      & EMS,   TCU, EPB, ICU, T-BOX, APA                \\
		0x68A & NVS         & 500     & HUD, PSW                                        \\
		\bottomrule 
	\end{tabular}}
	\label{table3}
\end{table}

Additionally, the frequency of the different IDs is fixed by the vehicle manufacturer. It is worth noting that important automotive components have a higher priority. Hence, time-series features are reflected in the ID of the CAN messages. Despite the fact that there are some event-triggered messages with variable frequency, the set of commands is also fixed. The details are shown in Table \ref{table3}.

Spatial feature expression, especially the data field of the CAN frame, is most significant. Most new messages are generated at a steady rate over the period of data acquisition in the dataset. In this study, the spatial signature analysis was based on the data segment, and a single message often had a regular change in timing. We analyzed this and aggregated the messages of different periodic variation rules. Table \ref{table4} displays some CAN message of different CAN IDs from the dataset, where the omissions represent same transferred bytes. We can observe that they have certain fixed byte constants (e.g., the first 7 bytes of ID=0x260), as well as fields that change more frequently (e.g., the third bytes of ID=0x316), but only in a certain range. There are distinct spatial characteristics of the data field, useful for modeling the normal behavior of the CAN packet, and also inspire the requirement to design attention convolutional blocks.

\begin{table}[t]
	\caption{Analyzing range statistics of message frame}
	\normalsize
	\centering
	\resizebox{\linewidth}{!}{
	\renewcommand{\arraystretch}{1.1} 
	\begin{tabular}{cccccccccc}%
		\toprule
		\textbf{ID}        & \textbf{Transmitter}     & \textbf{1}     & \textbf{2}     & \textbf{3}     & \textbf{4}     & \textbf{5}     & \textbf{6}     & \textbf{7}     & \textbf{8}     \\ 
		\midrule
		0x260     & N/A  & 05 & 22    & 00    & 30    & FF    & 99    & 63    & 38      \\
		&                 &            &       &       &       &       &       &      & 0B    \\
		&                 &            &       &       &       &       &       &       & 1A    \\
		&                 &            &       &       &       &       &       &       & 29    \\
		
		0x316     & RPM       & 05    & 22    & 6A    & 0B    & 21    & 18    & 00    & 7F    \\
		&                 &            &  22     & 16    &       & 22      &       &       &    \\
		&                 &            &  23     & 3A    &       &  23     &       &       &     \\
		&                 &            &  24     & 1A    &       &  24     &       &       &     \\
		
		0x43F     & GEAR        & 10    & 50    & 60    & FF    & 46    & 28    & 0A    & 00    \\
		&                 &            &       &       &       &       &  0C     &     &             \\
		&                 &            &       &       &       &       &    10   &    &           \\
		&                 &            &       &       &       &       &      F0 &     &             \\ 
		\bottomrule 
	\end{tabular}}
	\label{table4}
\end{table}

\begin{figure}[t]
	\centering	
	\includegraphics[width=1\linewidth]{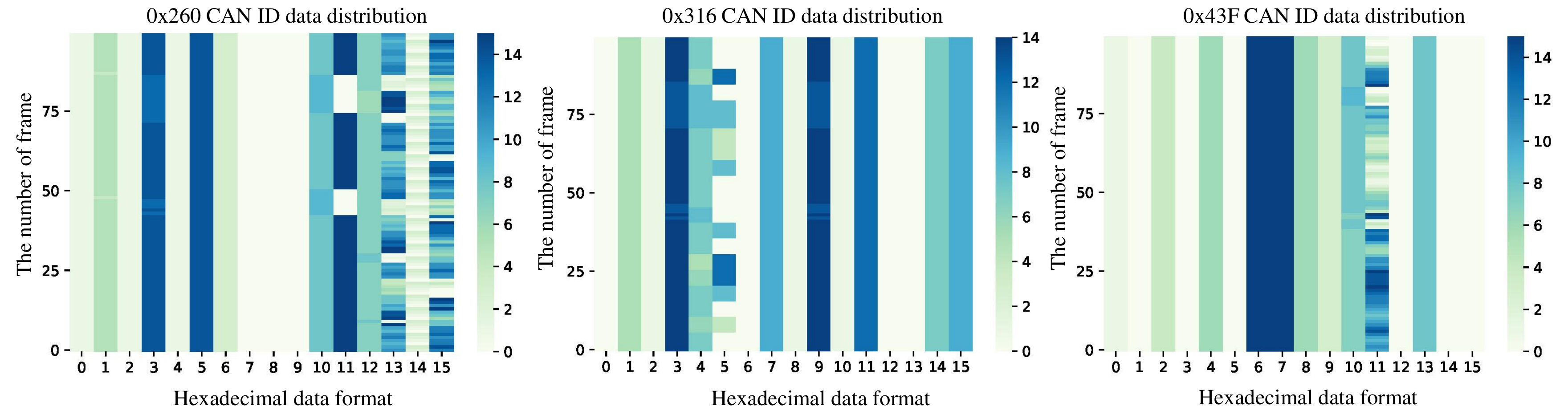}
	\caption{Value distribution by heatmap in hexadecimal form of each bit of CAN frame for different ID includes ID = 0x260, ID = 0x316, and ID=0x43F.}
	\label{Fig.5}
\end{figure}

To more visually observe the byte change patterns, the data fields for each message are displayed in a heat map, as shown in Fig \ref{Fig.5}. In order to apparently visualize the differences in variation of the each byte, this paper presents 100 consecutive communication messages based on three important CAN IDs such as gear and speed. There is a certain period of color shade variation at ID = 0x260 and ID = 0x43F, while the messages sent by the RPM-specific IDs are clearly irregular, corresponding to the summary in Table \ref{table4}.

In conclusion, spatial-temporal details are useful for modeling the normal behavior of CAN packets, and attention also focuses on bytes that change frequently and time-series important relationships, thus helping the model to quickly detect violations and determine the targeted traffic.

\section{IDS USING SPATIAL-TEMPORAL CORRELATION FEATURE}\label{sec4}
\subsection{Dataset preprocessing}
It is impractical to train an IDS based on the neural network on the original CAN dataset, so data pre-processing is a necessary part before model training. The payload field is 8 bytes as shown in Table \ref{table4}, where each byte is represented by two numbers in hexadecimal format. In fact, the public source dataset was progressively judged and found to contain a large number of data frames that were inferior to 8 bytes or irregular. To ensure uniformity of model input, we filled in the missing data frames with "00" in two scenarios: 1) where the data frame is less than 8 bytes; 2) where only the single digit "0" is used to represent a byte. Afterward, we split and transformed the arbitration bits and data domain in the dataset into a trainable dataset containing 19 features, respectively 16 features in 8 bytes of the data domain. In particular, the CAN ID is partitioned into 3 bits in order to harmonize the operation with the split data field. On the one hands, the combination of hexadecimal features of the original multi-bit will not be too far removed from the data field features due to the introduction of temporal-frequency features; on the other hand, all messages provided by the public dataset have only 3 valid bits in hexadecimal. 

In addition, the values of all features are converted to decimal from hexadecimal. After implementing missing data padding and decimal conversion to meet the basic inputs for the model, several additional data pre-processing steps still need to be completed. First, the CAN frame type is encoded using a label encoder, which is used to convert categorical features into numerical features owing to many ML\&DL-based algorithms cannot directly support string features \cite{yang2021mth}. Thereafter, the network dataset is normalized by the Min-Max algorithm, as the features collected in network traffic data often have a wide range of differences that impose model deviations and emphasize only large-scale features. Furthermore, the ML\&DL-based model is proven to perform more convergent easily on normalized dataset \cite{yang2020hyperparameter, RN02}. Hence, the data normalization by the Min-Max method is calculated as:

\begin{equation}
	X_{\text {norm }}=\frac{X-X_{\min }}{X_{\max }-X_{\min }}
\end{equation}

The method implements equal scaling of the original data, where $X_{\text {norm }}$ is the normalized data, $X$ is the original data, and $X_{\max}$ and $X_{\min}$ are the maximum and minimum values of the original dateset respectively. Table \ref{table5} presents the CAN data that is available as model input, where the former three columns are the CAN ID feature fields, the next eight columns in each message present 16 data domain fields, and the last column represents the label in digital form for each message. In addition, the first two rows represent the distribution for normal CAN data features, while the distribution for injection attacks is the third row. For multi-frame detection, encoder requires an additional image conversion step that splits the collected data set into $64 \times 19$ 2-D images. After the normalization completing, in order to prevent the same attacks from appearing in both the test and the segmented attack data, dataset division is for 10-fold cross-validation via StratifiedKFold function in sklearn library.

\begin{table}[t]
	\centering
	\normalsize
	\caption{The pre-processed dataset is represented with the labels divided into normal (0) and intrusion (1)}
	\resizebox{\linewidth}{!}{
		\renewcommand{\arraystretch}{1.3} 
	\begin{tabular}{cccccccccccc}%
		\toprule
		\multicolumn{3}{c}{ID}                                               & \multicolumn{8}{c}{DATA}                                                         & \multirow{2}{*}{Label} \\ \cmidrule{1-3}\cmidrule{4-11}
		ID1                   & ID2                   & ID3                  & Byte1 & Byte2 & Byte3 & Byte4 & Byte5 & Byte6 & Byte7 & Byte8                    &                        \\ \midrule
		\multirow{2}{*}{0.2}  & \multirow{2}{*}{0.06} & \multirow{2}{*}{0.4} & 0     & 0.13  & 0.4   & 0     & 0.13  & 0.13  & 0     & \multicolumn{1}{l}{0.4} & \multirow{2}{*}{0}     \\
		&                       &                      & 0.2   & 0.06  & 0.53  & 0.6   & 0.06  & 0.06  & 0     & \multicolumn{1}{l}{1}   &                        \\ \midrule
		\multirow{2}{*}{0.06} & \multirow{2}{*}{0.53} & \multirow{2}{*}{1}   & 1     & 0.33  & 0     & 0     & 0     & 0.2   & 0     & \multicolumn{1}{l}{0}   & \multirow{2}{*}{0}     \\
		&                       &                      & 0.93  & 0.73  & 0     & 0     & 0     & 0.8   & 0     & \multicolumn{1}{l}{0}   &                        \\ \midrule
		\multirow{2}{*}{0}    & \multirow{2}{*}{0}    & \multirow{2}{*}{0}   & 0     & 0     & 0     & 0     & 0     & 0     & 0     & \multicolumn{1}{l}{0}   & \multirow{2}{*}{1}     \\
		&                       &                      & 0     & 0     & 0     & 0     & 0     & 0     & 0     & \multicolumn{1}{l}{0}   &                        \\ 
		\bottomrule
	\end{tabular}}
	\label{table5}
\end{table}

\subsection{STC-IDS model design}
The STC-IDS consists of two steps: encoding and detection, respectively. The encoder is an analysis of the spatial-temporal characteristics of the CAN messages, and capturing the important relationships based on attention to it; the detector achieves anomaly classification to the valuable spatial-temporal features.

\subsubsection{STC-IDS for single-frame detection} 
From the perspective of single-frame intrusion detection, our aim is to retrieve the illegally controlled source ECU in conjunction with the CAN ID (i.e., accurate identification of every abnormal traffic). In training phase, we extract spatial features at one-dimensional data by CNN. Since the input of the proposed model is defined as $1\times19$, the spatial component extracts valuable features only through three convolutional blocks and a global pooling layer in order to avoid invalidating features because of the deeper network. Each convolutional block consists of a 1-D convolutional layer, a batch-normalization layer, and an activation function ReLU. The batch-normalization allows the model to discard the learning of biases, and make the convolutional output of the model homogeneous particularly. The ReLU function makes the network realize nonlinear feature mapping, while the global pooling serves to assist the model in searching for where critical bytes are and reducing redundant information.

Attention actually mimics a visual mechanism of the human brain, seen as an automatic weighting scheme \cite{jiang2022gatrust}. Hence, an improved LSTM structure with the attention mechanism (A-LSTM) is designed as a temporal component. We recognize the input as a multivariate time series with a single time step that is initially handled by the dimensional shuffling layer to increase the multivariate processing speed. When the features are fed into the A-LSTM blocks, it can mine significant temporal features. To prevent over-fitting, the discard layer plays a crucial role. Overall, the A-LSTM component could discover crucial byte changes, calculated as:

\begin{equation}
	a_{b}=\exp \left(u_{b}^{T} * u_{w}\right) / \sum_{b} \exp \left(u_{b}^{T} * u_{w}\right)
\end{equation}
Where $u_{w}$ is the weight matrix, and $u_{b}$ means the implicit representation of the hidden state $h_{b}$ at computation feature bit $b$. The $u_{b}$ is calculated as:
\begin{equation}
	u_{b}=\tanh \left(W_{w} h_{b}+b_{w}\right)
\end{equation}
Where $W_{w}$ is the weight matrix and $b_{w}$ is the bias. Afterward, we attain the attention probability distribution value at each byte. Finally, the final feature vector $v$ is calculated as:

\begin{equation}
	v=\sum_{b} a_{b} * h_{b}
\end{equation}

In Fig {Fig.6}, we present a parallel feature extraction classification model for single-frame detection. Both features are finally aggregated by the fully connected layers. We refer to it as the CAN valuable feature for spatial-temporal correlation under multi-view learning. Finally, it is fed into the final classification component, which specific elaboration in algorithm \ref{alg1}.

\begin{figure*}[t]
	\centering	
	\includegraphics[width=1\linewidth]{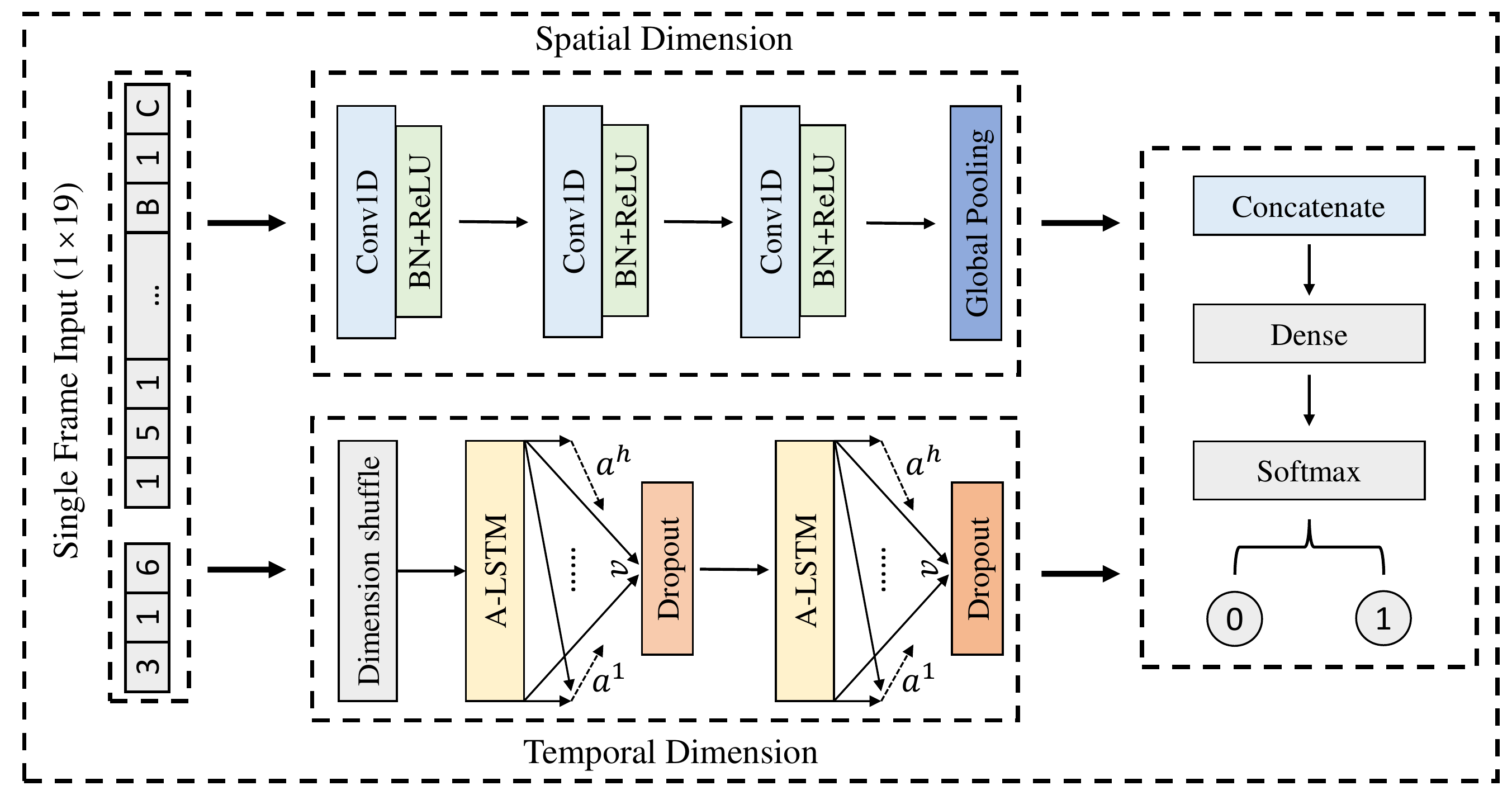}
	\caption{To illustrate the single-frame detection neural network.}
	\label{Fig.6}
\end{figure*}

\begin{algorithm}[t]
	\caption{STC-IDS for single-frame}
	\label{alg1}
	\begin{algorithmic}[1]
		\STATE Require: Input Data $X$, LSTM-Times = 2, Convolutional-Times = 3;
		\STATE Temporal Phase: 
		\STATE $X_{\text{temporal}}$ = Dimension shuffle($X$);
		\FOR{Times to LSTM-Times}
		\STATE $h_{b}$ = LSTM($X_{\text{temporal}}$);
		\STATE $u_{b}=\tanh \left(W_{w} h_{b}+b_{w}\right)$;
		\STATE $a_{b}=\exp \left(u_{b}^{T} * u_{w}\right) / \sum_{b} \exp \left(u_{b}^{T} * u_{w}\right)$;
		\STATE $v=\sum_{t} a_{b} * h_{b}$;
		\STATE $X_{\text{temporal}}$ = Dropout($v$);
		\ENDFOR
		\STATE Spatial Phase:
		\STATE $X_{\text{spatial}}$ = $X$;
		\FOR{Times to Convolutional-Times} 
		\STATE $X$ = Conv1D($X_{\text{spatial}}$)
		\STATE $X$ = Batch-Normalization($X$)
		\STATE $X$ = ReLU($X$)
		\STATE $X_{\text{spatial}}$ = $X$
		\ENDFOR
		\STATE $X_{\text{spatial}}$ = Global-Pooling($X_{\text{spatial}}$)
		\STATE $X_{\text{spatial-temporal}}$ = Concatenate($X_{\text{spatial}}$,$X_{\text{temporal}}$)
		\STATE $X_{\text{spatial-temporal}}$ = Dense($X_{\text{spatial-temporal}}$)
		\STATE $\hat{y}$ = Softmax($X_{\text{spatial-temporal}}$)
	\end{algorithmic}
\end{algorithm}

\subsubsection{STC-IDS for multi-frame detection}
To further elevate the efficiency, multi-frame based IDS is constructed. In other words, the model retrieves a continuous CAN 2-D matrix, aggregated from $64$ consecutive CAN messages during data pre-processing. The matrix height 64 represents the length of historical time series that the model could recall. For supervised learning, 2-D data frames that contain one or more injection messages are marked as attack CAN instance, while data frames that do not contain injection messages are marked as normal CAN instance. 

\begin{figure*}[t]
	\centering	
	\includegraphics[width=1\linewidth]{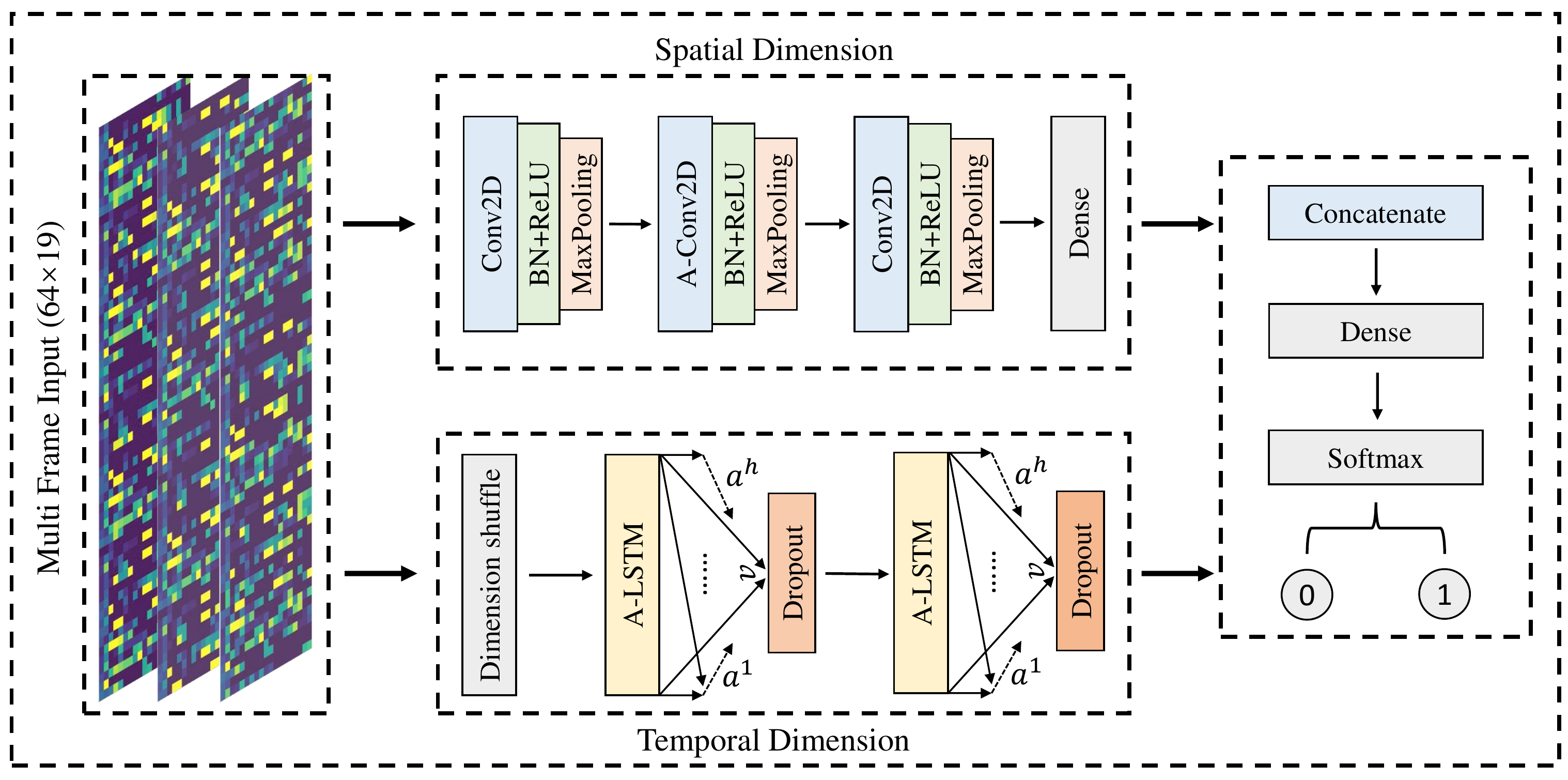}
	\caption{To illustrate the multi-frame detection neural network.}
	\label{Fig.7}
\end{figure*}

As shown in Fig \ref{Fig.7}, the 2-D data frames are fed into the parallel networks. For the temporal feature extraction component, we remains the structure as same as the single-frame model, but the model input is a time series in 2-D so that A-LSTM  can pay more attention to important relationships from previous time series. After the dimensional shuffling layer swap out the time dimension of the time series, the processed time-series are fed into the A-LSTM blocks. However, the three main parameters in the single-frame temporal attention model, i.e. $u_{b}$,$h_{b}$,$a_{b}$, need to be redefined as $u_{t}, h_{t}, a_{t}$. $u_{t}$ indicates the implicit representation of the hidden state $h_{t}$ at computation time $t$, while $a_{t}$ represents the final computed attention value.

Moreover, the spatial feature extraction component is inspired by the VGGNet model. we keep feature extraction capability of the model, but modify the number of convolutional layers and channel to adapt CAN dataset. It is composed of three convolutional blocks, where a convolutional module consists of a 2-D convolutional layer, a batch normalization layer, and a max-pooling layer. 

\begin{figure}[t]
	\centering	
	\includegraphics[width=1\linewidth]{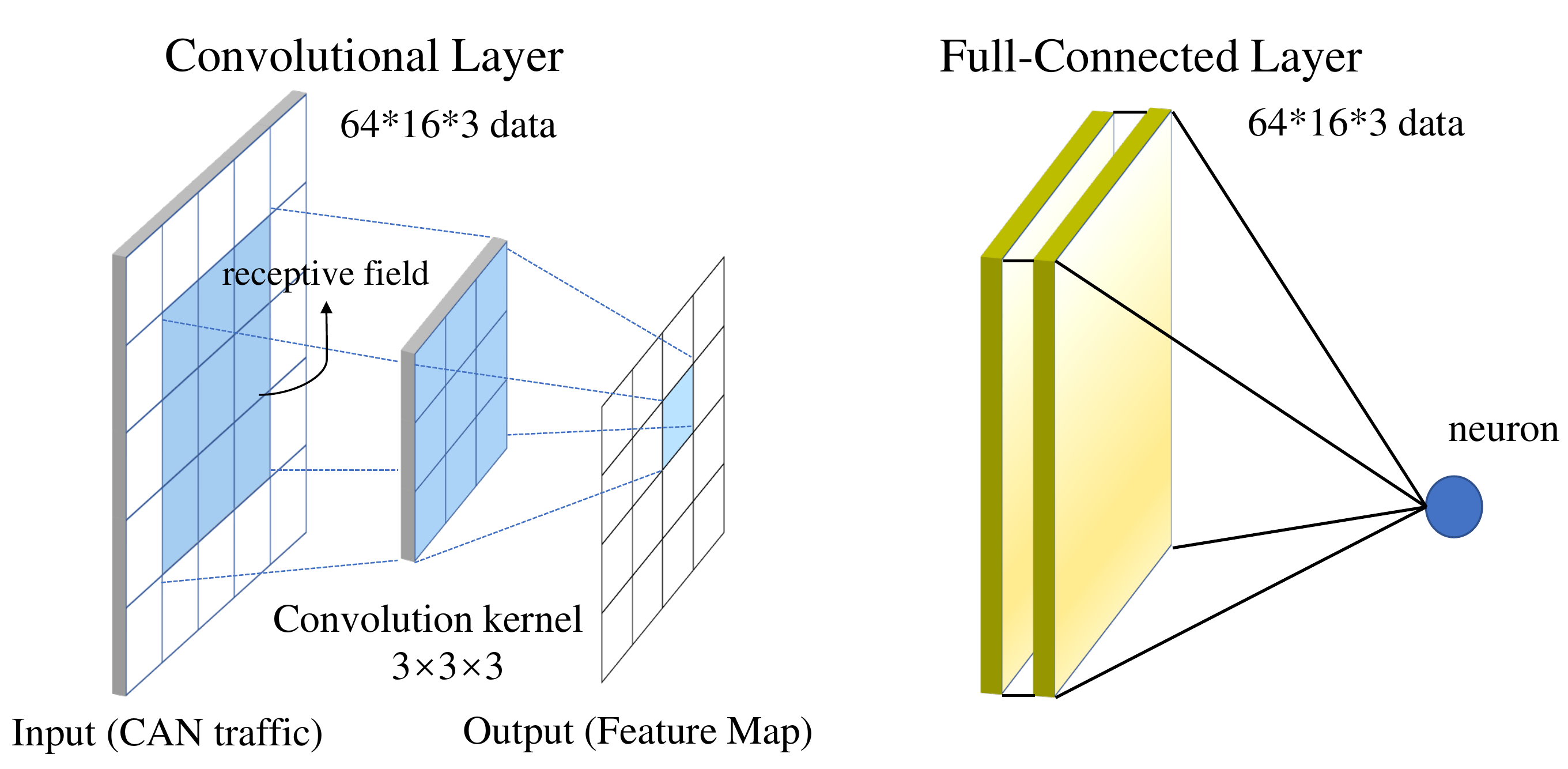}
	\caption{In the computation of neurons in the convolutional and fully connected layers, it is clear that the former is smaller than the latter, due to the dense connectivity.}
	\label{Fig.8}
\end{figure}

Since the convolutional layer has the feature of shared weight, the proposed model reduce the complexity and improve the inference efficiency. For instance, if a $64 \times 19 \times 3$ feature is mapped into a $62 \times 17 \times 6 $ volume, the full-connected layer requires $(64 \times 19 \times 3) \times (62 \times 17 \times 6) = 22M $ weights, while the convolutional layer via $3 \times 3$ convolutional kernels only require $(3\times3\times3) \times 6 = 162$ weights. Fig \ref{Fig.8} illustrates the difference between convolutional and fully connected layer computations. The multi-frame detection also has the capability to remove redundant information, and reduce the computational effort with the help of max-pooling layer. The convolutional block maps the raw data to the hidden feature space, thereby performing the task of feature engineering to improve the performance of the spatial component. The fully connected layer serves to map the learned "distributed feature representation" to the sample markup space. Hence, spatial-temporal correlation features are extracted in a parallel network, and then aggregated by a fully connected layer to finish the classification task.

Most importantly, the second convolutional component is enhanced with visual attention mechanism, called A-Conv2D. The core idea of the component is to help the network extract and represent the information most relevant to the target. For instance, we usually focus on significant information when viewing a photograph, and summarize it. Observing Table \ref{table2} and the heat map in Fig \ref{Fig.5}, there are clearly crucial information changes in the bytes as a 2-D data frames. Recently, channel and spatial attention are mentioned in the CNN \cite{woo2018cbam}. Inspired by the idea, the proposed structure integrates such ways, which can assist the model to find where the key information is and where the channel features are learned. The mode has the capability to self-select important features and eliminate the feature engineering step. Hence, convolutional branches of the intermediate layer can obtain a convolutional feature with spatial attention and channel attention. This additional structure is essentially cascaded over the original network with the purpose of better extracting valuable features. In summary, attention convolutional module features are calculated as follows:

\begin{equation}
	\begin{aligned}
		F_{\text {merge }} &=M_{s}\left(F_{c}\right) \otimes F_{c}\\
		&=M_{s}\left(M_{c}(F) \otimes F\right) \otimes\left(M_{c}(F) \otimes F\right)
	\end{aligned}
\end{equation}
where $F_{\text {merge }}$ is the aggregated feature. $M_{c}$, $M_{s}$ are channel attention weight coefficients and spatial attention wight coefficients, respectively, while $F$ is the input feature and $F_{c}$ is the channel attention feature.

\begin{figure*}[t]
	\centering	
	\includegraphics[width=1\linewidth]{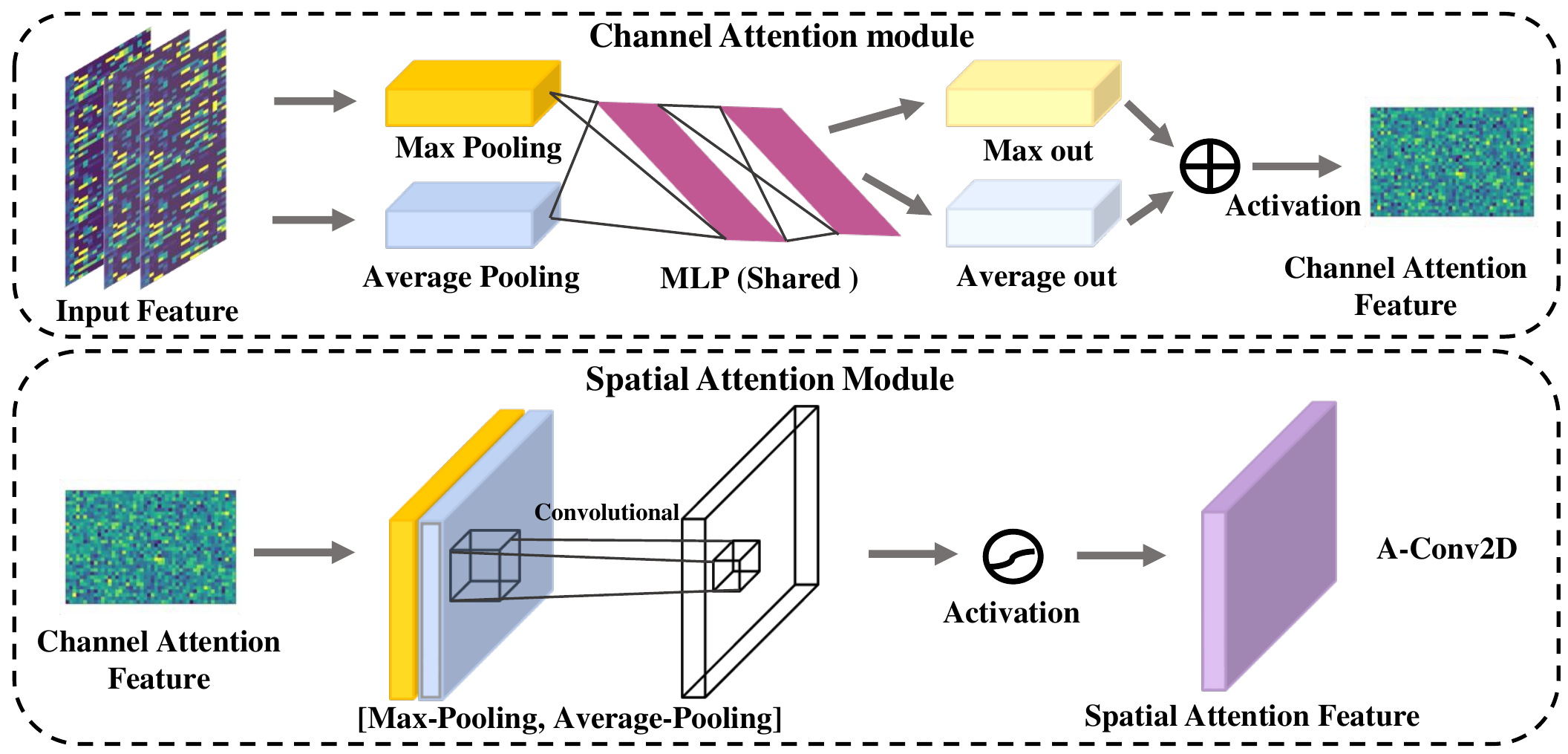}
	\caption{Schematic representation of the CAN image computed in the attention convolution module. The channel module utilizes the shared network to output the max-pooling and avg-pooling computed features; the spatial attention module pools the pooling output along the channel axis and computes the final features through the convolution layer.}
	\label{Fig.9}
\end{figure*}

In Fig \ref{Fig.9}, the feature maps built by the first convolution block are extracted deep features and attention coefficients through the attention convolution component. Initially, the max-pooling and avg-pooling layers aggregate spatial attention information to generate two different spatial context descriptions. Immediately afterward, the two features are added together by the multi-layer perceptron (MLP). Note that the MLP is weight-sharing to realize information sharing. Finally, the weight coefficients $M_{c}$ are obtained through the sigmoid activation function, which is calculated as: 
\begin{equation}
	\begin{aligned}
		M_{c}  &=\sigma(\mathrm{MLP}(\operatorname{AvgPool}(\mathrm{F}))+M L P(\text { MaxPool }(\mathrm{F}))) \\
		&=\sigma\left(\mathrm{W}_{1}\left(\mathrm{~W}_{0}\left(\mathrm{~F}_{\mathrm{avg}}^{\mathrm{c}}\right)\right)+\mathrm{W}_{1}\left(\mathrm{~W}_{0}\left(\mathrm{~F}_{\max }^{\mathrm{c}}\right)\right)\right)
	\end{aligned}
\end{equation}

After getting the weight coefficients $M_{c}$, the channel attention features $F_{c}$ is calculated as follows:
\begin{equation}
	F_{c}=M_{c}(F) \otimes F
\end{equation}
where $\sigma$ is the activation function, $W_{0}$ and $W_{1}$ are the weight matrices of the convergence layer.

The spatial attention module is a stable complement to channel attention information because of its ability to mine where the key features are. Similarly, given an $H \times W \times C$ feature $F_{c}$, two $H\times W \times 1$ channel descriptions are obtained by averaging pooling and max pooling in one channel dimension, respectively. Thereafter, the channel descriptions are stitched together based on the channel. Finally, a $2\times2$ convolutional layer with sigmoid activation function is applied to obtain the weight coefficients $M_{s}$, which calculated as follows:

\begin{equation}
	\begin{aligned}
		M_{\mathrm{s}} &=\sigma\left(f^{2 \times 2}\left(\left[\operatorname{AvgPool}\left(\mathrm{F}_{\mathrm{c}}\right) ; \operatorname{MaxPool}\left(\mathrm{F}_{\mathrm{c}}\right)\right]\right)\right) \\
		&=\sigma\left(f^{2 \times 2}\left(\left[\mathrm{~F}_{\text {avg }}^{s} ; \mathrm{F}_{\max }^{s}\right]\right)\right)
	\end{aligned}
\end{equation}
Where the $f^{2 \times 2}(\cdot)$ represents convolutional computation, $\mathrm{~F}_{\text {avg }}^{s}$ and $\mathrm{F}_{\max }^{s}$ represent two channel descriptions, respectively. Thus, the spatial-channel attention aggregation feature $F_{merge}$ is obtained by multiplying the weight coefficients $M_{s}$ with the feature $F_{c}$. The feature $F_{merge}$ is continued calculated by the normal convolutional component, and then aggregated with the temporal feature to build the spatial-temporal features. The fully connected layer maps the spatial-temporal feature to sample markup space to finish the two-class classification task. Specifically, the multi-frame model is described in algorithm \ref{alg2}

\begin{algorithm}[t]
	\caption{STC-IDS for multi-frame}
	\label{alg2}
	\begin{algorithmic}[1]
		\STATE Require: Input Data $X$, LSTM-Times = 2, Convolution-Times = 3;
		\STATE Temporal Phase: 
		\STATE $X_{\text{temporal}}$ = Dimension shuffle($X$);
		\FOR{Times to LSTM-Times}
		\STATE $h_{t}$ = LSTM($X_{\text{temporal}}$);
		\STATE $u_{t}=\tanh \left(W_{w} h_{t}+b_{w}\right)$;
		\STATE $a_{t}=\exp \left(u_{t}^{T} * u_{w}\right) / \sum_{t} \exp \left(u_{t}^{T} * u_{w}\right)$;
		\STATE $v=\sum_{t} a_{t} * h_{t}$;
		\STATE $X_{\text{temporal}}$ = Dropout($v$);
		\ENDFOR
		\STATE Spatial Phase:
		\STATE $X_{\text{spatial}}$ = $X$;
		\FOR{Times to Convolution-Times} 
		\STATE $F$ = Conv2D($X_{\text{spatial}}$)
		\STATE $F$ = Batch-Normalization($X$)
		\STATE $F$ = ReLU($F$)
		\STATE $F$ = Max-Pooling($F$)
		\IF{Times == 2}
		\STATE $M_{c} = $ Channel Attention($F$)
		\STATE $F_{c} = M_{c} \otimes F$
		\STATE $M_{s} = $ Spatial Attention($F_{c}$)
		\STATE $F_{merge} = M_{s} \otimes F_{c}$
		\STATE $X_{\text{spatial}}$ = $F_{merge}$
		\ELSE 
		\STATE $X_{\text{spatial}}$ = $F$
		\ENDIF
		\ENDFOR
		\STATE $X_{\text{spatial}}$ = Dense($X_{\text{spatial}}$)
		\STATE $X_{\text{spatial-temporal}}$ = Concatenate($X_{\text{spatial}}$,$X_{\text{temporal}}$)
		\STATE $X_{\text{spatial-temporal}}$ = Dense($X_{\text{spatial-temporal}}$)
		\STATE $\hat{y}$ = Softmax($X_{\text{spatial-temporal}}$)
	\end{algorithmic}
\end{algorithm}

\subsubsection{Classification loss}
For the intrusion detection based on classification, the loss value can be obtained according to the comparison between the prediction label and the actual label. Based on this loss message, the loss calculation method are used for the two data formats as discriminant mark for detection. Hence, the predict values $\hat{y}_{i}$ are calculated as follows. 
\begin{equation}
	\begin{aligned}
		\hat{y}_{i}&=p(c=i \mid x) \\
		&=\operatorname{softmax}\left(w \cdot f_{x}+b\right)(\text { for } \mathrm{i}=0,1)
	\end{aligned}
\end{equation}
Where $f_{x}$ is the combining features, and $\hat{y}_{i}$ present the probability distribution over target classes zero and one. With the fusion of two structures, the model becomes more complex generating over-fit phenomena easily. For better generalization of the model, $L2$ regularization is set in the network layer to limit the gradient. Besides, we need to continuously optimize by back propagation to reduce loss value, and fit the model to the best structure, in order to maximize the predicted probability $p$. The loss function is calculated as follows.
\begin{equation}
	\begin{aligned}
		L&=\frac{1}{N} \sum_{i} L_{i}+\lambda\|w\|^{2} \\
		&=\frac{1}{N} \sum_{i}-\left[y_{i} \cdot \log \left(p_{i}\right)+\left(1-y_{i}\right) \cdot \log \left(1-p_{i}\right)\right]+\lambda\|w\|^{2}
	\end{aligned}
\end{equation}

\section{Evaluation}\label{sec5}
We now validate that the spatial-temporal correlation feature can be used to detect anomaly frames, and evaluate the performance of a CAN bus prototype and real vehicles.

\subsection{Evaluation Metrics and Experiment Environment}
In this paper, statistical metrics TP (true positive) and TN (true negative) are introduced to indicate the number of frames correctly classified as attack and normal, while metrics FP (false positive) and FN (false negative) are introduced to indicate the number of data frames that are misclassified as attack and normal. The model accuracy formula is as follows.

\begin{equation}
	A c c=(T P+T N) /(T P+F N+F P+T N)
\end{equation}

Precision (P) and recall (R) are considered as evaluation metrics to assess classification performance. Precision refers to the rate at which the actual data frame labels are detected correctly, while recall represents the proportion of all attack frame samples that end up in the attack frame class, which are calculated as follows.

\begin{equation}
	P=T P /(T P+F P)
\end{equation}

and
\begin{equation}
	R=T P /(T P+F N)
\end{equation}

The F1 score evaluation is also presented, which is a harmonic average based on the detection precision and completeness. Also, the F1 score is often used to measure classification performance when the data are unevenly distributed, which is calculated as follows.

\begin{equation}
	F 1= {2 \times P \times R}/({P+R})
\end{equation}

Additionally, the false negative rate (FNR) and the error rate (ER) are one way of assessing classification performance. The FNR is the proportion of frames that are not detected as belonging to the attack frame and the ER is the proportion of frames that are incorrectly classified, calculated as follows.

\begin{equation}
	F N R=F N /(T P+F N)
\end{equation}

and
\begin{equation}
	E R=(F N+F P) /(T N+T P+F N+F P)
\end{equation}

The two models designed in this paper were trained offline based on the dataset, while the testing phase was based on real vehicles, injected with malicious frames of the same rules, to check the performance and efficiency of the models. The following is the experimental training and testing environment.

\begin{enumerate}
	\item Intel(R) Core (TM) i7-9500U CPU@3.6GHz
	\item RAM:64.0 GB
	\item GPU RTX 2080 Ti (Training environment)
	\item CAN Test, NVIDIA Jetson AGX Xavier (16GB) (Testing environment)
\end{enumerate}

\subsection{Hyperparameter Selection and Optimization}\label{sec5.2}
The selection of hyperparameters is a crucial step in network performance and inference efficiency. Currently, automated hyperparametric optimization (HPO) services and tool-kits address the constant trial-and-error steps of deep learning developers \cite{yu2020hyper}. In this paper, bayesian optimization (BO) automatic parametric tuning is used in order to quickly determine hyperparameters. It is a typical method applied to global optimization problems. Compared to grid search and stochastic search, BO is more computationally efficient and requires fewer attempts to find the optimal set of hyperparameters \cite{yang2020hyperparameter}.

Based on the BO optimization library provided by Keras-tuner and on a 10-fold cross-validation dataset, we selected important hyperparameters such as learning rate (1e-2, 1e-3, 1e-4, 1e-5, 1e-6), optimizer (e.g., Adam, SGD, RMSprop), dropout rate, and filters. Afterward, we set the optimization goal to validate the accuracy (Val-Acc), a maximum number of trials of 10, and train the model 3 times per trial. Finally, the optimization results will present the set of top 3 hyperparameters for performance.

\begin{table}[t]
	\caption{Effect of hyperparameters on model performance under automated HPO}
	\centering
	\resizebox{\linewidth}{!}{
		\renewcommand{\arraystretch}{1.3}
	\begin{tabular}{cccccccccc}%
		\toprule
		Model   type                    & Learning   rate & Dropout   rate & Filters            & Dense        & Optimizer     & Val-Acc         \\ \midrule
		\multirow{3}{*}{Single   Frame} & \textbf{1e-6}   & \textbf{0.4}   & \textbf{16,32,128} & \textbf{64}  & \textbf{Adam} & \textbf{0.9998} \\
		& 1e-6            & 0.4            & 8,96,32            & 48           & Adam          & 0.9935          \\
		& 1e-6            & 0.4            & 8,16,192           & 80           & Adam          & 0.9869          \\ \midrule
		\multirow{3}{*}{Multi   Frame}  & \textbf{1e-2}   & \textbf{0.4}   & \textbf{64,16,32}  & \textbf{128} & \textbf{Adam} & \textbf{0.9996} \\
		& 1e-2            & 0              & 8,16,32            & 128          & Adam          & 0.9993          \\
		& 1e-2            & 0              & 24,80,96           & 256          & Adam          & 0.9992          \\ 
		\bottomrule	
	\end{tabular}}
	\label{table6}
\end{table}

As shown in Table \ref{table6}, the average accuracy on the cross-validation set reaches 99.98\% in the single-frame model when setting the learning rate is 1e-6, the dropout rate is 0.4, the filters are 16, 32, 128, the dense layer is 64, and the optimizer is Adam. Similarly, the accuracy is achieved up to 99.96\% to the multi-frame model under the optimal hyperparameters.

\begin{figure}[t]
	\centering	
	\includegraphics[width=1\linewidth]{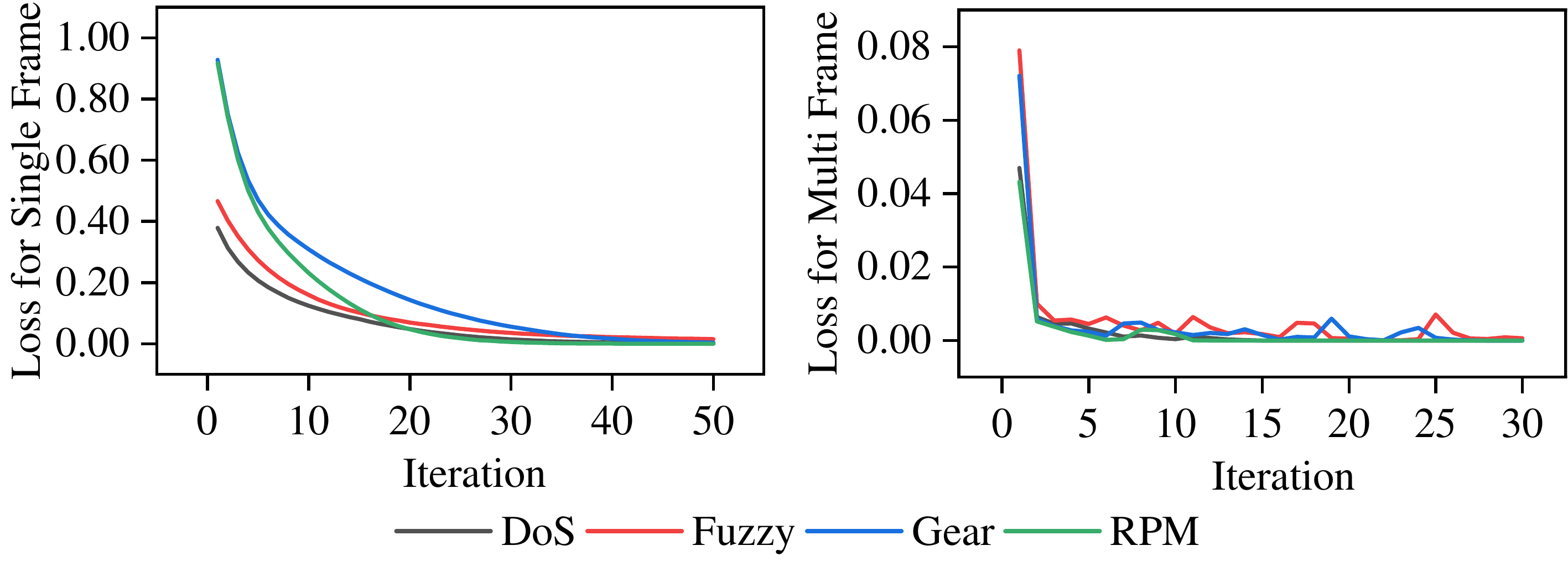}
	\caption{To illustrate training loss on each dataset in single-frame (a) and multi-frame (b) algorithms.}
	\label{Fig.10}
\end{figure}

Interestingly, the inference speed of the model was significantly accelerated aided by the optimal hyperparameters. Despite the higher detection performance was obtained by the single-frame model, the inference speed is slower. Relative to multi-frame, it took 50 iterations to converge, whereas the multi-frame model only took 30 generations. Fig \ref{Fig.10} shows the iterative training losses for each dataset. It can be observed that since the DoS attack disrupts the frequency of injection, both models learn the patterns and converge quickly. However, as the complexity of attack data increases, the fuzzy and spoofing attacks injected with random data converge slower in the single-frame model, while the training loss convergence fluctuates significantly in the multi-frame model.

\subsection{Detection Metrics Evaluation and Comparison}

\begin{figure*}[t]
	\centering	
	\includegraphics[width=1\linewidth]{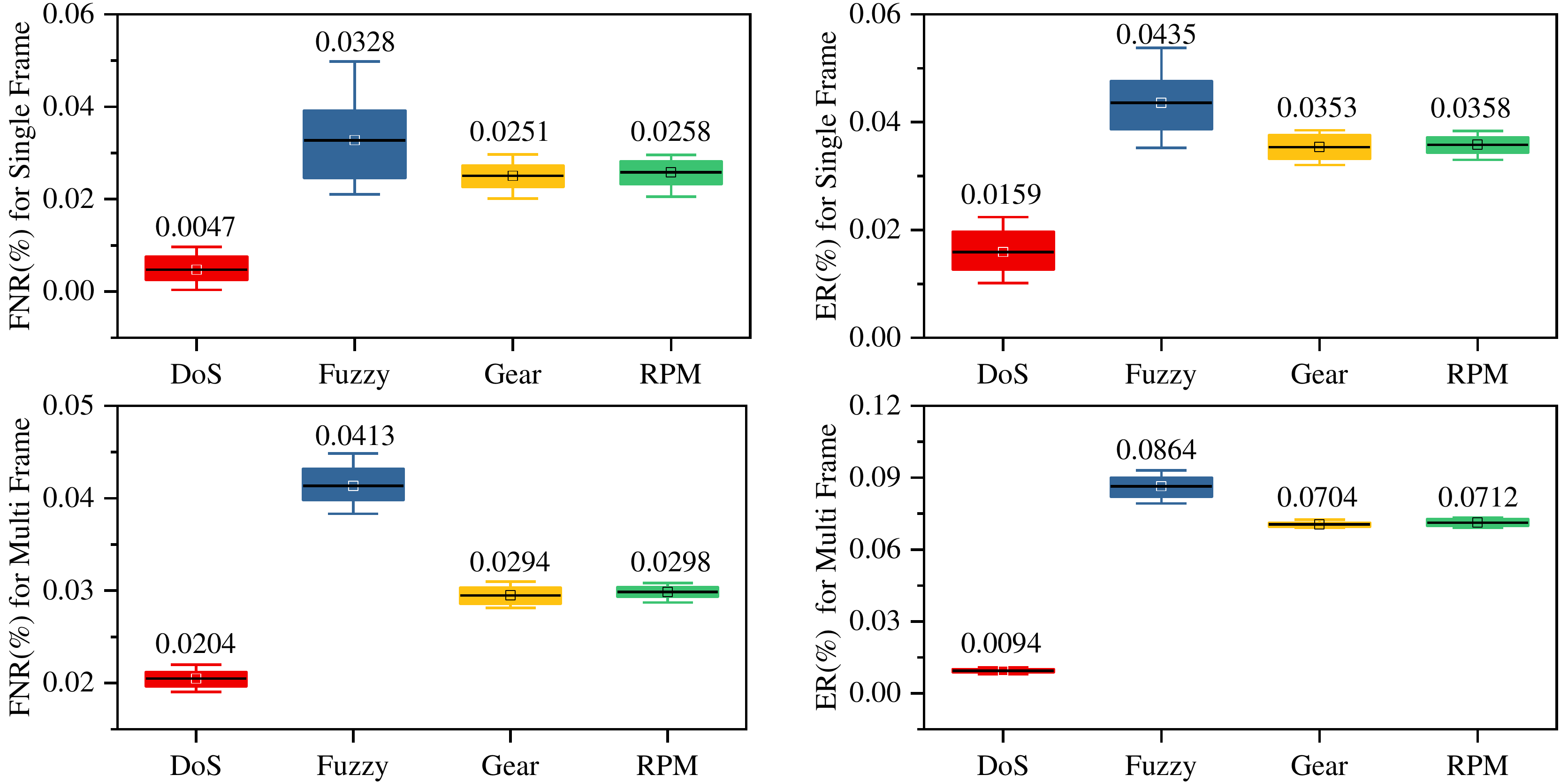}
	\caption{Box-plots of single-frame and multi-frame models measuring FNR and ER in 30 replicate experiments.}
	\label{Fig.11}
\end{figure*}

After training the best two models according to the set hyperparameters, Fig \ref{Fig.11} shows the detection performance of the models in terms of ER and FNR for the four-attack testing sets in 30 repeated experiments. The two proposed models, both in terms of FNR, and ER, presents stable performance to detect DoS attacks with mean values of 0.0047\%, 0.0204\%, 0.0159\% and 0.0094\%, respectively. On the contrary, the fuzzy attacks shows greater fluctuations in detection performance. The complexity of fuzzy attack data seems to be significantly higher than the other injected data, necessitating more training iterations to construct a stable model. Hence, the ultimate mean values of FNR are still 0.0328\% and 0.0413\%, as well as ER get 0.0435\% and 0.0864\%. 

Although FNR and ER on spoofing attacks are higher than DoS attacks, they achieve stable and better results compared to fuzzy attacks. The single-frame detector averages 0.0251\% and 0.0353\% for the Gear attack on both metrics, while the multi-frame detector obtains 0.0294\% and 0.0704\%. Similarly, the performance of both models is similar to Gear attacks when detecting RPM attacks. Notably, the single-frame detection model requires mining the characteristics of different CAN packets, thus exhibiting fluctuations in FNR and ER values that are greater than the multi-frame model. Despite the multi-frame model with better detection efficiency, it has a higher false alarm rate than the former.

To reflect the advantages of our model, Table \ref{table7} lists the detection performance of STC-IDS when compared to those of the other machine-learning techniques, where the highest performance values are highlighted in bold, and "-" only means that the scheme has not been tested on this metric. The results indicate clearly that the STC-IDS model outperforms the previous methods on all datasets. The false negative rate and error rate are significantly reduced. It can be seen that the model captures spatial-temporal details of network traffic remarkably well and enhances the anomaly detection ability satisfactorily.

\begin{table}[t]
	\caption{IDS performance comparsion with baseline methods}
	\centering
	\resizebox{\linewidth}{!}{
		\renewcommand{\arraystretch}{1.3}
	\begin{tabular}{lccccccccc}%
		\toprule
		DoS & ER (\%) & FNR (\%) & P (\%) & R (\%) & F1 (\%) \\ \midrule
		\textbf{STC-IDS   for Single Frame} & 0.02 & \textbf{0.01}  & 99.95 & \textbf{99.99} & \textbf{99.96} \\
		\textbf{STC-IDS   for Multi Frame}  & \textbf{0.01} & 0.02   & 99.91 & \textbf{99.97} & 99.94 \\
		3-LSTM \cite{pawelec2019towards}                     & 0.07 & 0.22   & \textbf{1.0}   & 99.78 & 99.88 \\
		DCNN  \cite{song2020vehicle}                     & 0.03 & 0.10   & \textbf{1.0}   & 99.89 & 99.95 \\
		DAE    \cite{amarbayasgalan2018unsupervised}                    & -    & 0.12   & 91.27 & 99.88 & 95.36 \\
		OTIDS     \cite{lee2017otids}                 & -    & 26.2 & 99.90 & 73.80 & 84.88 \\ \midrule	Fuzzy                      & ER (\%)       & FNR (\%)      & P (\%)         & R (\%)         & F1 (\%)        \\ \midrule
		\textbf{STC-IDS   for Single Frame} & \textbf{0.05}           & \textbf{0.04} & \textbf{99.97} & \textbf{99.95} & \textbf{99.96} \\ 
		\textbf{STC-IDS   for Multi Frame}  & 0.09 & 0.07          & 99.90          & 99.92          & 99.91          \\
		3-LSTM \cite{pawelec2019towards}                  & 0.84          & 0.65          & 99.36          & 99.16          & 99.26          \\
		DCNN  \cite{song2020vehicle}                      & 0.18          & 0.35          & 99.95          & 99.65          & 99.80          \\
		DAE   \cite{amarbayasgalan2018unsupervised}                     & -             & 3.74          & 90.05          & 96.26          & 93.05          \\
		OTIDS   \cite{lee2017otids}                   & -             & 29.79         & 99.14          & 70.21          & 82.20     \\ \midrule
		Gear                                & ER (\%)       & FNR (\%)      & P (\%)         & R (\%)         & F1 (\%)        \\ \midrule
		\textbf{STC-IDS   for Single Frame} & \textbf{0.04}          & \textbf{0.03} & 99.97          & \textbf{99.96} & \textbf{99.97} \\
		\textbf{STC-IDS   for Multi Frame}  & 0.07 & \textbf{0.03} & 99.94          & \textbf{99.96} & 99.95          \\
		3-LSTM \cite{pawelec2019towards}                            & 0.24          & 0.32          & 99.75          & 99.68          & 99.72          \\
		DCNN  \cite{song2020vehicle}                             & 0.05          & 0.11          & \textbf{99.99} & 99.89          & 99.94          \\
		DAE     \cite{amarbayasgalan2018unsupervised}                            & -             & 18.2          & 94.63          & 81.80          & 87.75          \\
		OTIDS  \cite{lee2017otids}                            & -             & 28.35         & 99.83          & 71.65          & 83.42    \\ \midrule
		RPM                                & ER (\%)       & FNR (\%)      & P (\%)     & R (\%)         & F1 (\%)        \\ \midrule
		\textbf{STC-IDS   for Single Frame} & 0.04          & \textbf{0.03} & 99.98      & \textbf{99.96} & \textbf{99.97} \\
		\textbf{STC-IDS   for Multi Frame}  & 0.07 & \textbf{0.03} & 99.95      & \textbf{99.96} & 99.96          \\
		3-LSTM \cite{pawelec2019towards}            & 0.13          & 0.30          & \textbf{1} & 99.71          & 99.85          \\
		DCNN  \cite{song2020vehicle}    & \textbf{0.03}          & 0.05          & 99.99      & 99.94          & 99.96          \\
		DAE                              \cite{amarbayasgalan2018unsupervised}   & -             & 4.27          & 95.73      & 95.73          & 92.10          \\
		OTIDS                   \cite{lee2017otids}            & -             & 28.32         & 99.81      & 71.68          & 83.43         \\
		\bottomrule
	\end{tabular}}
	\label{table7}
\end{table}

Evidently, we can observe that the STC-IDS for single-frame model achieves a stable precision (99.97\%), a higher recall (99.96\%), and an outstanding F1-score (99.96\%) in average as compared to a threshold (OTIDS), classification (DCNN), prediction (3-LSTM), and clustering (DAE) based models. In contrast, the performance solely decreases by an average of 0.04\% precision, 0.01\% recall, and 0.03\% in the F1-score while maintaining efficiency in the multi-frame model.

The 3-LSTM scheme shows high accuracy and unstable recall because of the unconsidered spatial correlation, resulting in a low F1 score and a large gap between FNR and ER with our scheme and DCNN. DCNN is currently one of the best models for in-vehicle intrusion detection, but the stacking of convolutions is ineffective in capturing the temporal relationships of the inputs. As a result, the single-frame model reduces the FNR and ER by 90\% and 33\% over the DCNN scheme for DoS attacks, while the multi-frame model reduces the FNR by 80\% and the ER by 66\%. More notably, for complex fuzzy attacks, only 0.04\% FNR and 0.05\% ER are achieved on the single-frame model. In addition, the FNR is slightly improved on the Gear and RPM dataset. Meanwhile, the ER gets approximate performance compared with the DCNN model. 

Compared to the DAE model with a lack of labeling constraints, the proposed model has larger improvements in detection performance, while DAE only reaches 95.36\% in F1-score on DoS attacks, and obtains lower performance on other attacks. Similarly, the OTIDS exhibited a high accuracy rate and extremely low recall, resulting in an FNR of over 25\%, reflecting a relatively low F1-score.  Thus, the robust spatial-temporal correlation feature demonstrates sufficient benefits in terms of improving detection performance and reducing the false alarm rate.

\subsection{Time Cost in Real Vehicle}
\begin{figure}[t]
	\centering	
	\includegraphics[width=1\linewidth]{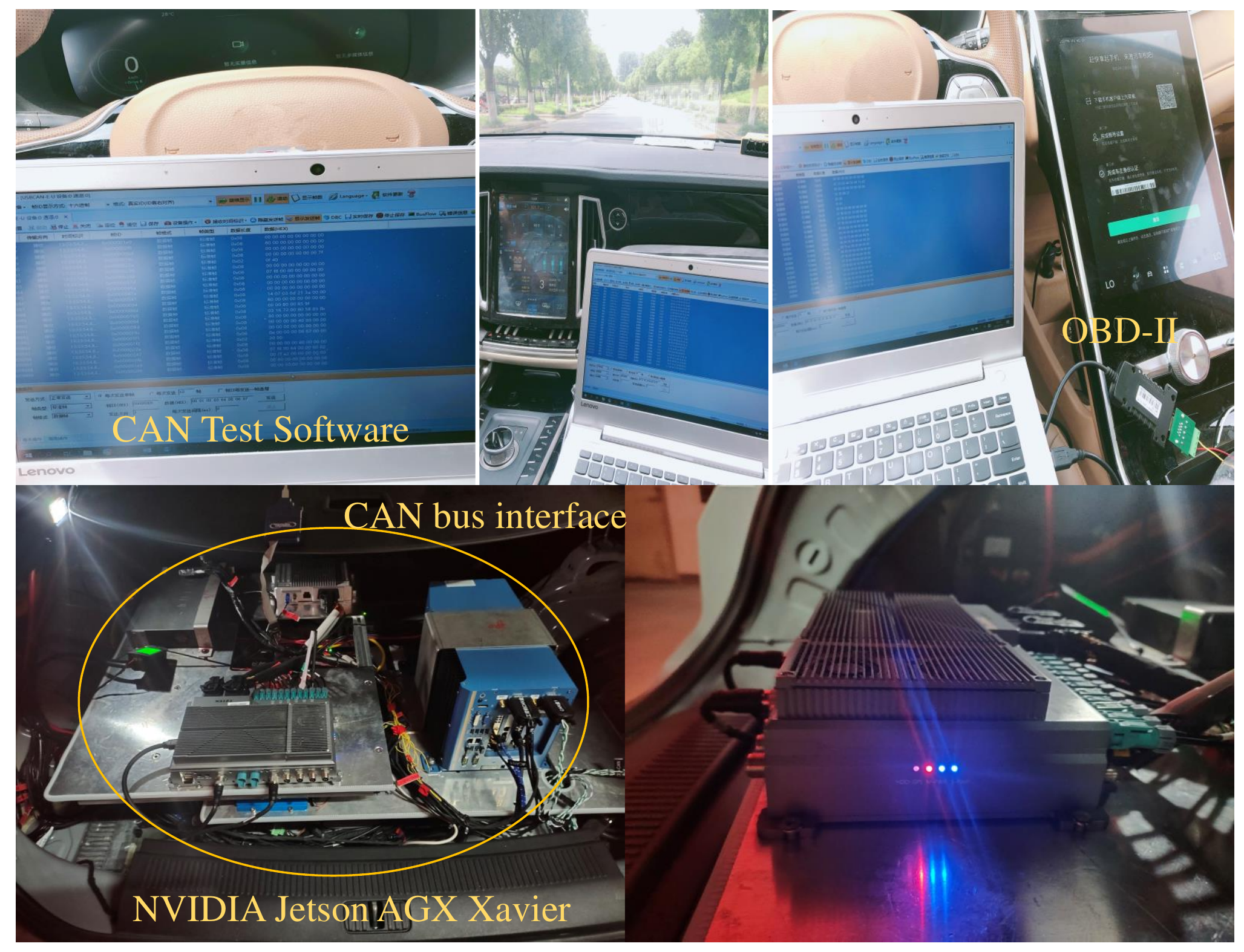}
	\caption{To illustrate the testing experiment environment.}
	\label{Fig.12}
\end{figure}

This work first implemented single frame detection in terms of security considerations. Although the performance is improved over previous algorithms, the efficiency is not guaranteed.  Besides, time and resources are the major limitations to applying deep learning models to real-world vehicle IDS. Therefore, STC-IDS based on multi-frames also was proposed. Obviously, significant improvements were made in terms of model convergence time as shown in subsection \ref{sec5.2}. To test the efficiency of proposed model a resource-constrained in-vehicle network, the model was tested on an in-vehicle class device the NVIDIA Jetson AGX Xavier. Note that only 4GB of video memory was allocated for testing. The CAN Test software injected the attack traffic by connecting to the OBD-II port, as shown in Fig \ref{Fig.12}.

\begin{figure}[t]
	\centering	
	\includegraphics[width=1\linewidth]{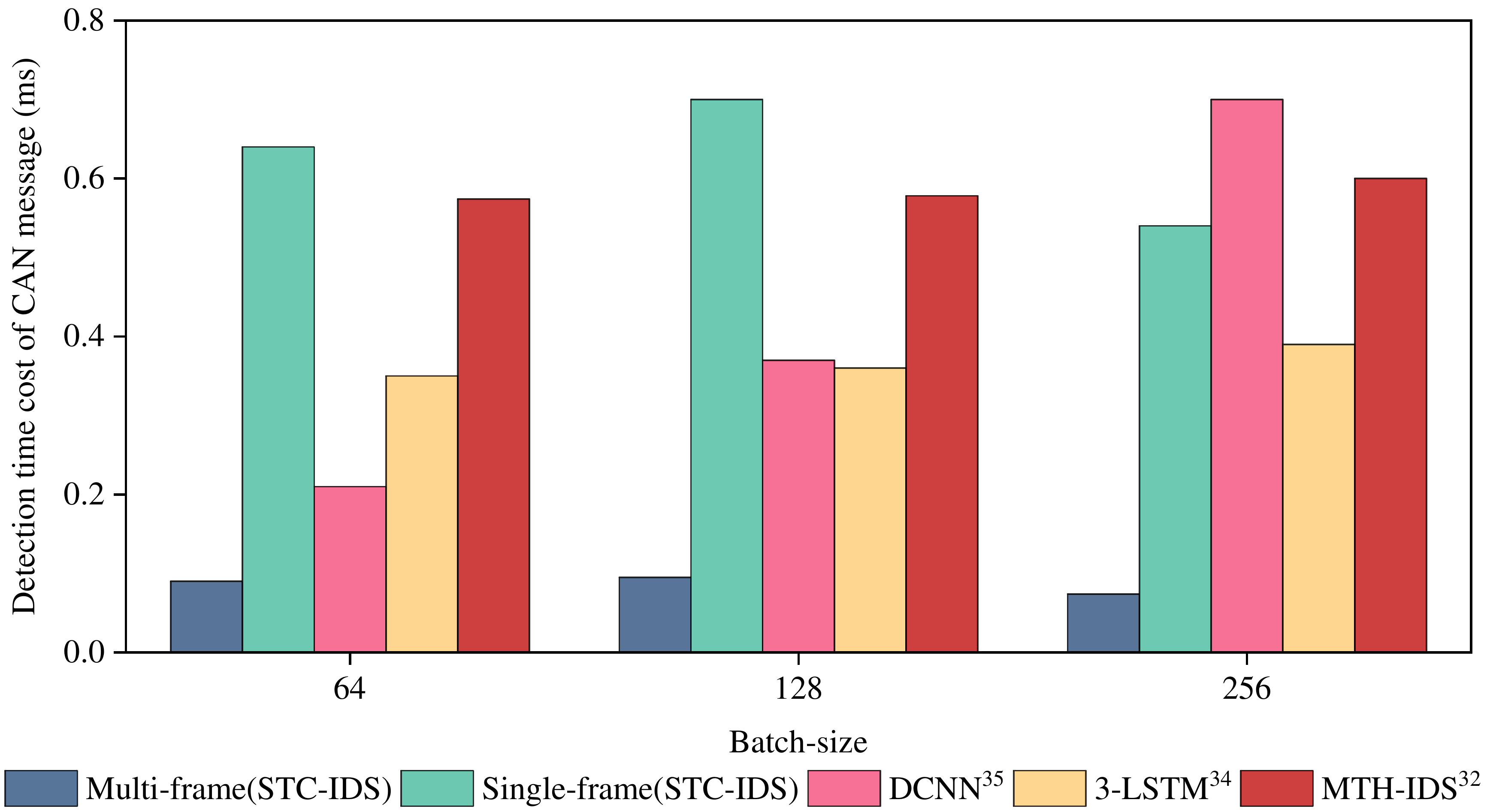}
	\caption{Testing time of proposed models on different batches for each message compared to other algorithms.}
	\label{Fig.13}
\end{figure}

In Fig \ref{Fig.13}, we present that the proposed model has lower time (in milliseconds) cost on anomaly detection in case of different batches compared to previous algorithms. The average detection time for single-frame model remains under 0.7 ms. Although small-batch requires the highest time cost, the detection time is decreased based on the increasingly batch. For example, 256 batch shows 0.54 ms time cost, which is superior to DCNN \cite{song2020vehicle} and MTH-IDS \cite{yang2021mth}. Moreover, this truly indicates that the model is constantly learning as it reasoned, thus making it easier to catch malicious features later on. However, it can be estimated that the model can detect 1851 messages in 1s  at the fastest detection time cost, while CAN frames transmit approximately 2000 messages in 1s. One limitation of this method is that they do not usually satisfy real-time detection.

To overcome this limitation, the multi-frame model presents prominent advantages that just need the average time cost of 0.09 ms, 0.09 ms, and 0.074 ms for three batches, respectively. Compared testing time of the previous models as listed in Fig \ref{Fig.13}, the proposed model not only present outstanding performance, but also improves about 5 times in terms of efficiency. This means that the model can infer about five times the number of CAN transmission messages in 1s. Hence, the proposed model has the feasibility for real-time detection. Unlike the DCNN \cite{song2020vehicle} model, a crucial phenomenon is that the proposed model does not depends on batch size. The same result reflects on 3-LSTM model \cite{pawelec2019towards} and MTH-IDS \cite{yang2021mth}. But even so, we also suggest a suitable batch size needs to be determined, as large batches may delay anomaly alert.

\subsection{Discussion and Limitations}
The study presents two enhanced spatial-temporal features analyzing IDS for detecting single CAN messages and consecutive CAN frames based on open injection attack datasets. Experiments indicates that the single-frame model exhibits good detection performance. Similarly, the STC-IDS for multi-frame improves efficiency while ensuring performance.

Although the detection efficiency of the single-frame model is right limited, we believe that the model will be suitable for real-time detection when using higher-performance computing devices. In fact, it is necessary to track unauthorized ECUs in conjunction with CAN ID indexing in further in-vehicle security development based on single-frame detection. It is worth noting that the time-cost detection of proposed methods is based on in-vehicle edge computing and autonomous driving platform from our research group. Although it has satisfied the requirement for real-time intrusion detection when in a real environment,  it has some impact when all tasks are performed simultaneously. Hence, more research still needs to develop in terms of practical implementation. 

Clearly, the spatial-temporal feature modeling in this study is based on observing the CAN ID domain, data field, and communication protocol of the specific brand vehicles. The generality consequence of the proposed model is demonstrated due to the fixed time-interval, sender and receiver, despite different vehicle companies making distinct communication protocols. In other words, model transfer only needs to be re-trained on the new brand vehicle, which also can extract valuable spatial-temporal features, and obtains excellent detection performance.

Moreover, the division of CAN ID in the scheme is limited to public datasets. In order to accommodate more realistic in-vehicle messages, the division should be completed in practice by considering both standard frames (11 bits) and extended frames (29 bits).  However, it is straightforward only to modify the dimensionality of the input for the proposed model. Most importantly, the model has a fundamental limitation in terms of detecting unlearned types of attacks as it is based on supervised learning. To address this challenge, more research is needed on unknown attack detection using models with generative functions such as adversarial training, or auto-encoder.

\section{Conclusions}
This work focuses on learning the temporal and spatial characteristics of in-vehicle network traffic in order to establish enhanced spatial-temporal correlation features, and then build an automotive intrusion detection model. The proposed model is implemented based on the encode-detection architecture. The encoding layer is constructed as a parallel network in both temporal and spatial terms base on LSTM and CNN. The introduction of A-LSTM and A-Conv2D helps the model uncover important relationships between keyword node variation and temporal order. Spatial-temporal correlation features are fed into the detection layer to complete anomaly detection.

Both models achieve optimal hyperparameter selection with Bayesian optimization, reducing the number of iterations and elevating accuracy. Compared to previous schemes, our model achieves a better performance on open source dataset, i.e., DoS, Fuzzy, Gear, and RPM, especially in the FNR and ER metrics.

Although this study achieves security protection of the CAN bus, there is still much room for improvement, especially unknown attack detection. In future work, we will consider how to express realistic unknown attack messages, improving model robustness and generalization capabilities. Data annotation also is a tedious task, but unsupervised algorithms are one of the solutions, which still need continuous research to improve performance.
\bibliographystyle{IEEEtran} 

\bibliography{Manuscript}

\begin{IEEEbiography}[{\includegraphics[width=1in,height=1.25in]{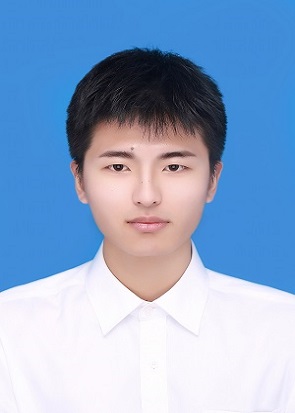}}]{Pengzhou Cheng} received the M.S. Degree with the Department of Computer Science and Communication Engineering, Jiangsu University, Zhenjiang, China, in 2022. He is currently pursuing the Ph.D. Degree with the Department of Electronic Information and Electrical Engineering, Shanghai Jiao Tong University, Shanghai, 201100, China. His primary research interests include cybersecurity, machine learning, time-series data analytic, intelligent transportation systems, and intrusion detection system.
\end{IEEEbiography}

\vspace{-10mm}

\begin{IEEEbiography}[{\includegraphics[width=1in,height=1.25in,clip,keepaspectratio]{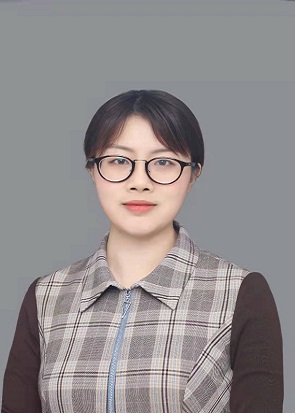}}]{Mu Han} (Member, IEEE) received the Ph.D.degree in Computer Science from Nanjing University of Science and Technology. She is an Associate Professor with the School of Computer Science and Communication Engineering, Jiangsu University. Her primary research interests are in the areas of Cryptography, Security and Communication in-vehicle networks, the design of security protocols for smart cars and Information Security.
\end{IEEEbiography}

\begin{IEEEbiography}[{\includegraphics[width=1in,height=1.25in]{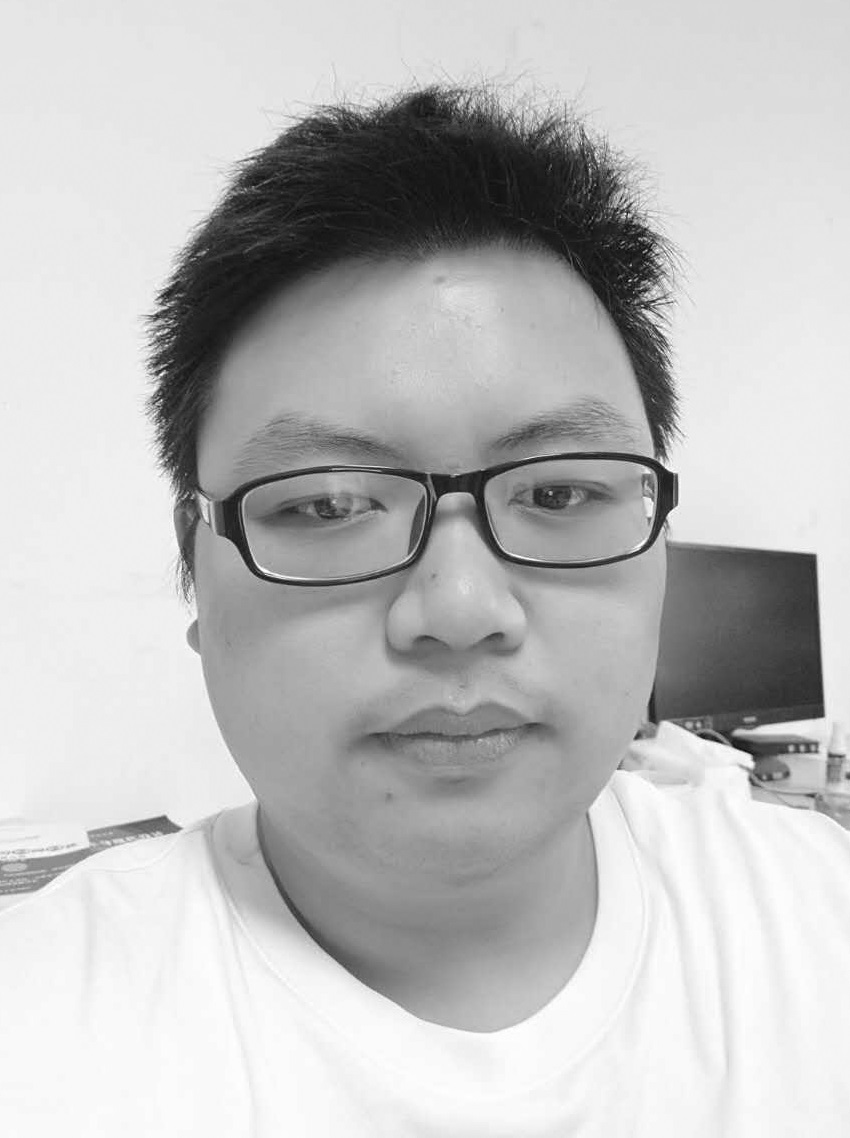}}]{Aoxue Li} received the B.S., M.S., and Ph.D. degrees in vehicle engineering from Jiangsu University, Zhenjiang, China, in 2013, 2016, and 2020, respectively. From August 2018 to August 2019, he was a visiting scholar with the Department of Mechanical Engineering, Michigan State University, East Lansing, MI, USA. He is currently a lecturer with the School of Automotive and Traffic Engineering, Jiangsu University. His research interests include autonomous vehicle, intelligent transportation systems, and ADAS technologies.
\end{IEEEbiography}

\begin{IEEEbiography}[{\includegraphics[width=1 in,height=1.25in]{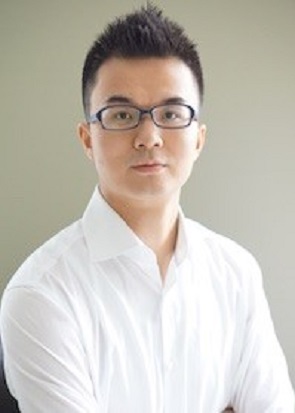}}]{Fengwei Zhang} received the Ph.D.degree in Computer Science from George Mason University. He is an Associate Professor with Department of Computer Science and Engineering, Southern University of Science and Technology. He was an Assistant Professor and the Director of the COMPASS Lab with Department of Computer Science, Wayne State University. His primary research interests are in the areas of systems security, with a focus on trustworthy execution, hardware-supported security, transparent malware analysis, and plausible deniability encryption.
\end{IEEEbiography}

\end{document}